\documentclass[aps,prd,twocolumn,showkeys,amsmath,amssymb]{revtex4}
\usepackage{graphicx}
\usepackage{epstopdf}   
\usepackage{multirow}
\usepackage{subfigure}
\usepackage[colorlinks,citecolor=blue,anchorcolor=red,menucolor=red,linkcolor=red,filecolor=red,runcolor=red,urlcolor=blue,frenchlinks=red]{hyperref}

\allowdisplaybreaks[3]

\begin{document}

\title{Decay properties of the $X(3872)$ through the Fierz rearrangement}

\author{Hua-Xing Chen}
\email{hxchen@seu.edu.cn}
\affiliation{
School of Physics, Southeast University, Nanjing 210094, China
}

\begin{abstract}
We systematically construct all the tetraquark currents of $J^{PC} = 1^{++}$ with the quark configurations $[cq][\bar c \bar q]$, $[\bar c q][\bar q c]$, and $[\bar c c][\bar q q]$ ($q=u/d$). Their relations are derived using the Fierz rearrangement of the Dirac and color indices, through which we study decay properties of the $X(3872)$ under both the compact tetraquark and hadronic molecule interpretations. We propose to search for the $X(3872) \rightarrow \chi_{c0} \pi$, $\eta_c \pi \pi$, and $\chi_{c1} \pi \pi$ decay processes in particle experiments.
\end{abstract}

\maketitle

\section{Introduction}
\label{sec:intro}

Since the discovery of the $X(3872)$ in 2003 by Belle~\cite{Choi:2003ue}, lots of charmonium-like $XYZ$ states were discovered in the past twenty years~\cite{pdg}. They are good candidates of four-quark states consisting of two quarks and two antiquarks, and their experimental and theoretical studies have significantly improved our understanding of the strong interaction at the low energy region. Although there is still a long way to fully understand how the strong interaction binds these quarks and antiquarks together with gluons, this subject has become and will continuously be one of the most intriguing research topics in hadron physics~\cite{Liu:2019zoy,Lebed:2016hpi,Esposito:2016noz,Guo:2017jvc,Ali:2017jda,Olsen:2017bmm,Karliner:2017qhf,Brambilla:2019esw}.

The $X(3872)$ is the most puzzling state among all the charmonium-like $XYZ$ states. Although it is denoted as the $\chi_{c1}(3872)$ in PDG2018~\cite{pdg}, the mass of the charmonium state $\chi_{c1}(2P)$ was estimated to be $3.95$~GeV~\cite{Godfrey:1985xj}, significantly higher than the $X(3872)$. This challenges the interpretation of the $X(3872)$ as a conventional charmonium state, and various interpretations were proposed to explain it, such as a compact tetraquark state composed of a diquark and an antidiquark~\cite{Maiani:2004vq,Maiani:2014aja,Hogaasen:2005jv,Ebert:2005nc,Barnea:2006sd,Chiu:2006hd}, a loosely-bound hadronic molecular state composed of two charmed mesons~\cite{Voloshin:1976ap,Tornqvist:2004qy,Close:2003sg,Voloshin:2003nt,Wong:2003xk,Braaten:2003he,Swanson:2003tb}, and a hybrid charmonium state with constituents $c \bar cg$~\cite{Close:2003mb,Li:2004sta}, etc. There were also some studies of the $X(3872)$ as a conventional $c \bar c$ state~\cite{Barnes:2003vb,Eichten:2004uh,Quigg:2004vf,Kong:2006ni}, and it was considered as the mixture of a  $c \bar c$ state with a $D \bar D^{*}$ component in Refs.~\cite{Meng:2005er,Meng:2014ota,Suzuki:2005ha}. We refer to reviews~\cite{Liu:2019zoy,Lebed:2016hpi,Esposito:2016noz,Guo:2017jvc,Ali:2017jda,Olsen:2017bmm,Karliner:2017qhf,Brambilla:2019esw} for detailed discussions.

The charged charmonium-like state $X(3872)$ of $J^{PC}=1^{++}$~\cite{Aaij:2015eva} has been observed in the $D^{0} \bar D^{*0}$, $J/\psi \pi \pi$, $J/\psi \omega$, and $\gamma J/\psi$ decay channels~\cite{Gokhroo:2006bt,Aubert:2007rva,Adachi:2008sua,Choi:2011fc,Abe:2005ix,delAmoSanchez:2010jr,Ablikim:2019zio,Aubert:2008ae,Bhardwaj:2011dj,Aaij:2014ala}, and there have been some evidences for the $X(3872) \to \gamma \psi(2S)$ decay~\cite{Aubert:2008ae,Aaij:2014ala}. Especially, its decay channels $J/\psi \pi \pi$ and $J/\psi \omega$ have comparable branching ratios~\cite{Abe:2005ix,delAmoSanchez:2010jr,Choi:2011fc,Ablikim:2019zio}, implying a large isospin violation. In a recent BESIII experiment~\cite{Ablikim:2019soz}, evidence for the $X(3872) \to \chi_{c1} \pi$ decay was reported with a statistical significance of $5.2\sigma$ using data at center-of-mass energies between 4.15 and 4.30 GeV, while this was not confirmed in the later Belle experiment~\cite{Bhardwaj:2019spn}. We refer to a recent paper~\cite{Li:2019kpj}, where the authors presented a complete analysis of all the existing experimental data and determine the absolute branching fractions of the $X(3872)$ decays. We also refer to another recent paper~\cite{Braaten:2019ags}, which studies branching fractions of the $X(3872)$ from a theoretical point of view. There have been many experimental and theoretical studies on this subject, and we refer to reviews~\cite{Liu:2019zoy,Lebed:2016hpi,Esposito:2016noz,Guo:2017jvc,Ali:2017jda,Olsen:2017bmm,Karliner:2017qhf,Brambilla:2019esw} for more discussions.

In Ref.~\cite{Chen:2019wjd} we have studied decay properties of the $Z_c(3900)$ through the Fierz rearrangement of the Dirac and color indices, and in this paper we shall apply the same method to study decay properties of the $X(3872)$. Both of these two studies are based on our previous finding that the diquark-antidiquark currents ($[qq][\bar q \bar q]$) and the meson-meson currents ($[\bar q q][\bar q q]$) are related to each other through the Fierz rearrangement of the Dirac and color indices~\cite{Chen:2006hy,Chen:2006zh,Chen:2007xr,Chen:2013jra}. More studies on light baryon operators can be found in Refs.~\cite{Chen:2008qv,Chen:2009sf}. The present study follows the idea of the QCD factorization method~\cite{Beneke:1999br,Beneke:2000ry,Beneke:2001ev}, which has been widely and successfully applied to study weak decay properties of (heavy) hadrons.

The $X(3872)$, as either a compact tetraquark state or a hadronic molecular state, contains four quarks. There can be three configurations ($q=u/d$):
\begin{equation*}
[cq][\bar c \bar q]\, , ~~ [\bar c q][\bar q c]\, , ~~ {\rm and} ~~ [\bar c c][\bar q q] \, .
\end{equation*}
In this paper we shall apply the Fierz rearrangement to relate them, and extract some strong decay properties of the $X(3872)$ under both the compact tetraquark and hadronic molecule interpretations. We shall not calculate the absolute values of these decay widths, but extract their relative branching ratios, which are also useful to understand the nature of the $X(3872)$~\cite{Yu:2017zst}. A similar arrangement in the nonrelativistic case was used to study strong decay properties of the $X(3872)$ and $Z_c(3900)$ in Refs.~\cite{Voloshin:2013dpa,Maiani:2017kyi,Voloshin:2018pqn}.

This paper is organized as follows. In Sec.~\ref{sec:current} we systematically construct all the tetraquark currents of $J^{PC}=1^{++}$ with the quark content $c \bar c q \bar q$. We consider three configurations, $[cq][\bar c \bar q]$, $[\bar c q][\bar q c]$, and $[\bar c c][\bar q q]$, and we derive their relations using the Fierz rearrangement of the Dirac and color indices. In Sec.~\ref{sec:decaydiquark} and Sec.~\ref{sec:decaymolecule} we extract some isoscalar decay channels of the $X(3872)$, separately for the compact tetraquark and hadronic molecule interpretations, and in Sec.~\ref{sec:isospin} we investigate its isovector decay channels. The obtained results are discussed in Sec.~\ref{sec:summary}, and formulae of decay amplitudes and decay widths are given in \ref{sec:width}.

\section{Tetraquark currents of $J^{PC} = 1^{++}$ and their relations}
\label{sec:current}

\begin{figure*}[hbt]
\begin{center}
\subfigure[~\mbox{$[cq][\bar c \bar q]$ currents $\eta_\mu^i(x,y)$}]{\includegraphics[width=0.23\textwidth]{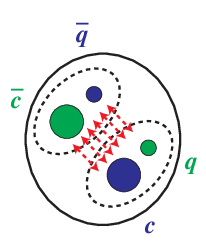}}
~~~~~
\subfigure[~\mbox{$[\bar c q][\bar q c]$ currents $\xi_\mu^i(x,y)$}]{\includegraphics[width=0.23\textwidth]{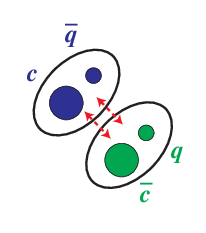}}
~~~~~
\subfigure[~\mbox{$[\bar c c][\bar q q]$ currents $\theta_\mu^i(x,y)$}]{\includegraphics[width=0.23\textwidth]{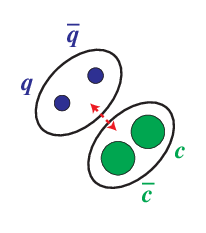}}
\caption{Three types of tetraquark currents. Quarks and antiquarks are shown in red, green, and blue color.}
\label{fig:current}
\end{center}
\end{figure*}

Similar to Ref.~\cite{Chen:2019wjd}, we can use the $c$, $\bar c$, $q$, $\bar q$ quarks ($q=u/d$) to construct three types of tetraquark currents of $J^{PC} = 1^{++}$, as illustrated in Fig.~\ref{fig:current}:
\begin{eqnarray}
\nonumber    \eta(x,y)   &=& [q^{\rm T}_a(x)~\mathbb{C} \Gamma_1~c_b(x)]   \times   [\bar q_c(y)~\Gamma_2 \mathbb{C}~\bar c_d^{\rm T}(y)] \, ,
\\           \xi(x,y)    &=& [\bar c_a(x)~\Gamma_3~q_b(x)]           \times   [\bar q_c(y)~\Gamma_4~c_d(y)] \, ,
\\ \nonumber \theta(x,y) &=& [\bar c_a(x)~\Gamma_5~c_b(x)]           \times   [\bar q_c(y)~\Gamma_6~q_d(y)] \, ,
\end{eqnarray}
where $\Gamma_i$ are Dirac matrices, and the subscripts $a, b, c, d$ are color indices. We separately construct them in the following subsections.

Generally speaking, one can apply the Fierz rearrangement to relate the local diquark-antidiquark currents $\eta(x,x)$ and the local meson-meson currents $\xi(x,x)$ and $\theta(x,x)$, but this equivalence is just between diquark-antidiquark and mesonic-mesonic currents, while compact diquark-antidiquark tetraquark states and weakly-bound meson-meson molecular states are totally different. To exactly describe them, one needs to explicitly use non-local currents to perform QCD sum rule analyses, but we are still not able to do this.

%
%

\subsection{$[qc][\bar q \bar c]$ currents $\eta_\mu^i(x,y)$}
\label{sec:current1}

There are eight independent $[qc][\bar q \bar c]$ currents of $J^{PC} = 1^{++}$~\cite{Chen:2010ze}:
\begin{eqnarray}
\eta^1_\mu &=& q_a^{\rm T}\mathbb{C}\gamma_\mu c_b ~ \bar q_{a} \gamma_5\mathbb{C}\bar c_{b}^{\rm T} + q_a^{\rm T}\mathbb{C}\gamma_5 c_b ~ \bar q_{a} \gamma_\mu\mathbb{C}\bar c_{b}^{\rm T} \, ,
\\
\nonumber \eta^2_\mu &=& q_a^{\rm T}\mathbb{C}\gamma_\mu c_b ~ \bar q_{b} \gamma_5\mathbb{C}\bar c_{a}^{\rm T} + q_a^{\rm T}\mathbb{C}\gamma_5 c_b ~ \bar q_{b} \gamma_\mu\mathbb{C}\bar c_{a}^{\rm T} \, ,
\\
\nonumber \eta^3_\mu &=& q_a^{\rm T}\mathbb{C}\gamma^\nu c_b ~ \bar q_{a} \sigma_{\mu\nu} \gamma_5\mathbb{C}\bar c_{b}^{\rm T} + q_a^{\rm T}\mathbb{C}\sigma_{\mu\nu} \gamma_5 c_b ~ \bar q_{a} \gamma^\nu\mathbb{C}\bar c_{b}^{\rm T} \, ,
\\
\nonumber \eta^4_\mu &=& q_a^{\rm T}\mathbb{C}\gamma^\nu c_b ~ \bar q_{b} \sigma_{\mu\nu} \gamma_5\mathbb{C}\bar c_{a}^{\rm T} + q_a^{\rm T}\mathbb{C}\sigma_{\mu\nu} \gamma_5 c_b ~ \bar q_{b} \gamma^\nu\mathbb{C}\bar c_{a}^{\rm T} \, ,
\\
\nonumber \eta^5_\mu &=& q_a^{\rm T}\mathbb{C}\gamma_\mu \gamma_5 c_b ~ \bar q_{a}\mathbb{C}\bar c_{b}^{\rm T} + q_a^{\rm T}\mathbb{C}c_b ~ \bar q_{a} \gamma_\mu \gamma_5\mathbb{C}\bar c_{b}^{\rm T} \, ,
\\
\nonumber \eta^6_\mu &=& q_a^{\rm T}\mathbb{C}\gamma_\mu \gamma_5 c_b ~ \bar q_{b}\mathbb{C}\bar c_{a}^{\rm T} + q_a^{\rm T}\mathbb{C}c_b ~ \bar q_{b} \gamma_\mu \gamma_5\mathbb{C}\bar c_{a}^{\rm T} \, ,
\\
\nonumber \eta^7_\mu &=& q_a^{\rm T}\mathbb{C}\gamma^\nu \gamma_5 c_b ~ \bar q_{a} \sigma_{\mu\nu}\mathbb{C}\bar c_{b}^{\rm T} + q_a^{\rm T}\mathbb{C}\sigma_{\mu\nu} c_b ~ \bar q_{a} \gamma^\nu \gamma_5\mathbb{C}\bar c_{b}^{\rm T} \, ,
\\
\nonumber \eta^8_\mu &=& q_a^{\rm T}\mathbb{C}\gamma^\nu \gamma_5 c_b ~ \bar q_{b} \sigma_{\mu\nu}\mathbb{C}\bar c_{a}^{\rm T} + q_a^{\rm T}\mathbb{C}\sigma_{\mu\nu} c_b ~ \bar q_{b} \gamma^\nu \gamma_5\mathbb{C}\bar c_{a}^{\rm T} \, .
\end{eqnarray}
In the above expressions we have omitted the coordinates $x$ and $y$. The color structures of $\eta^1_\mu - \eta^2_\mu$, $\eta^3_\mu - \eta^4_\mu$, $\eta^5_\mu - \eta^6_\mu$, and $\eta^7_\mu - \eta^8_\mu$ are all antisymmetric $[q c]_{\mathbf{\bar 3}_c}[\bar q \bar c]_{\mathbf{3}_c}\rightarrow[q c \bar q \bar c]_{\mathbf{1}_c}$, and those of $\eta^1_\mu + \eta^2_\mu$, $\eta^3_\mu + \eta^4_\mu$, $\eta^5_\mu + \eta^6_\mu$, and $\eta^7_\mu + \eta^8_\mu$ are all symmetric $[q c]_{\mathbf{6}_c}[\bar q \bar c]_{\mathbf{\bar 6}_c}\rightarrow[q c \bar q \bar c]_{\mathbf{1}_c}$.

In the diquark-antidiquark model proposed in Refs.~\cite{Maiani:2004vq,Maiani:2014aja,Shi:2021jyr}, the authors use $|s_{qc}, s_{\bar q \bar c} \rangle_J$ to denote ground-state tetraquarks, where $s_{qc}$ and $s_{\bar q \bar c}$ are the charmed diquark and antidiquark spins, respectively. They interpret the $X(3872)$ as a compact tetraquark state of $J^{PC} = 1^{++}$:
\begin{equation}
|0_{qc}1_{\bar q \bar c}; 1^{++} \rangle = {1\over\sqrt2} \left(| 0_{qc}, 1_{\bar q \bar c} \rangle_{J=1} + |1_{qc}, 0_{\bar q \bar c} \rangle_{J=1} \right) \, .
\label{def:diquark}
\end{equation}
The interpolating current having the identical internal structure is the current $\eta^1_\mu - \eta^2_\mu$, which has been studied in Refs.~\cite{Navarra:2006nd,Matheus:2006xi,Chen:2010ze,Wang:2013vex,Azizi:2017ubq,Padmanath:2015era}:
\begin{eqnarray}
\eta^{\mathcal X}_\mu(x,y) &=& \eta^1_\mu(x,y) - \eta^2_\mu(x,y)
\label{current:diquark}
\\ \nonumber &=& q_a^{\rm T}(x)\mathbb{C}\gamma_\mu c_b(x)  \left( \bar q_{a}(y) \gamma_5\mathbb{C}\bar c_{b}^{\rm T}(y) - \{ a \leftrightarrow b \} \right)
\\ \nonumber && +  \{ \gamma_\mu \leftrightarrow \gamma_5 \} \, .
\end{eqnarray}

\subsection{$[\bar c q][\bar q c]$ currents $\xi_\mu^i(x,y)$}
\label{sec:current2}

There are eight independent $[\bar c q][\bar q c]$ currents of $J^{PC} = 1^{++}$:
\begin{eqnarray}
\xi^1_\mu &=& \bar c_{a} \gamma_\mu q_a ~ \bar q_{b} \gamma_5 c_b - \bar c_{a} \gamma_5 q_a ~ \bar q_{b} \gamma_\mu c_b \, ,
\\
\nonumber \xi^2_\mu &=& \bar c_{a} \gamma^\nu q_a ~ \bar q_{b} \sigma_{\mu\nu} \gamma_5 c_b + \bar c_{a} \sigma_{\mu\nu} \gamma_5 q_a ~ \bar q_{b} \gamma^\nu c_b \, ,
\\
\nonumber \xi^3_\mu &=& \bar c_{a} \gamma_\mu \gamma_5 q_a ~ \bar q_{b} c_b + \bar c_{a} q_a ~ \bar q_{b} \gamma_\mu \gamma_5 c_b \, ,
\\
\nonumber \xi^4_\mu &=& \bar c_{a} \gamma^\nu \gamma_5 q_a ~ \bar q_{b} \sigma_{\mu\nu} c_b - \bar c_{a} \sigma_{\mu\nu} q_a ~ \bar q_{b} \gamma^\nu \gamma_5 c_b \, ,
\\
\nonumber \xi^5_\mu &=& {\lambda^n_{ab}}{\lambda^n_{cd}} \times \left( \bar c_{a} \gamma_\mu q_b ~ \bar q_{c} \gamma_5 c_d - \bar c_{a} \gamma_5 q_b ~ \bar q_{c} \gamma_\mu c_d \right) \, ,
\\
\nonumber \xi^6_\mu &=& {\lambda^n_{ab}}{\lambda^n_{cd}} \times \left( \bar c_{a} \gamma^\nu q_b ~ \bar q_{c} \sigma_{\mu\nu} \gamma_5 c_d + \bar c_{a} \sigma_{\mu\nu} \gamma_5 q_b ~ \bar q_{c} \gamma^\nu c_d \right) \, ,
\\
\nonumber \xi^7_\mu &=& {\lambda^n_{ab}}{\lambda^n_{cd}} \times \left( \bar c_{a} \gamma_\mu \gamma_5 q_b ~ \bar q_{c} c_d + \bar c_{a} q_b ~ \bar q_{c} \gamma_\mu \gamma_5 c_d \right) \, ,
\\
\nonumber \xi^8_\mu &=& {\lambda^n_{ab}}{\lambda^n_{cd}} \times \left( \bar c_{a} \gamma^\nu \gamma_5 q_b ~ \bar q_{c} \sigma_{\mu\nu} c_d - \bar c_{a} \sigma_{\mu\nu} q_b ~ \bar q_{c} \gamma^\nu \gamma_5 c_d \right) \, .
\end{eqnarray}
The former four $\xi^{1,2,3,4}_\mu$ have the color structure $[\bar c q]_{\mathbf{1}_c}[\bar q c]_{\mathbf{1}_c}\rightarrow[c \bar c q \bar q]_{\mathbf{1}_c}$, and the latter four $\xi^{5,6,7,8}_\mu$ have the color structure $[\bar c q]_{\mathbf{8}_c}[\bar q c]_{\mathbf{8}_c}\rightarrow[c \bar c q \bar q]_{\mathbf{1}_c}$.

In the molecular picture the $X(3872)$ is interpreted as the $D \bar D^*$ hadronic molecular state of $J^{PC} = 1^{++}$~\cite{Voloshin:1976ap,Tornqvist:2004qy,Close:2003sg,Voloshin:2003nt,Wong:2003xk,Braaten:2003he,Swanson:2003tb}:
\begin{equation}
| D \bar D^*; 1^{++} \rangle = {1\over\sqrt2} \left(| D \bar D^* \rangle_{J=1} + | \bar D D^* \rangle_{J=1} \right)  \, ,
\label{def:molecule}
\end{equation}
and the relevant interpolating current is just $\xi^1_\mu(x,y)$, which has been studied in Refs.~\cite{Lee:2008uy,Zhang:2009em,Wang:2013daa,Prelovsek:2013cra}:
\begin{eqnarray}
\xi^{\mathcal X}_\mu(x,y) &=& \xi^1_\mu(x,y)
\label{current:molecule}
\\ \nonumber &=& \bar c_{a}(x) \gamma_\mu q_a(x) ~ \bar q_{b}(y) \gamma_5 c_b(y) - \{ \gamma_\mu \leftrightarrow \gamma_5 \} \, .
\end{eqnarray}

\subsection{$[\bar c c][\bar q q]$ currents $\theta_\mu^i(x,y)$}
\label{sec:current3}

There are eight independent $[\bar c c][\bar q q]$ currents of $J^{PC} = 1^{++}$:
\begin{eqnarray}
\theta^1_\mu(x,y) &=& \bar c_{a}(x) c_a(x) ~ \bar q_{b}(y) \gamma_\mu \gamma_5 q_b(y) \, ,
\\
\nonumber \theta^2_\mu(x,y) &=& \bar c_{a}(x) \gamma_\mu \gamma_5 c_a(x) ~ \bar q_{b}(y) q_b(y) \, ,
\\
\nonumber \theta^3_\mu(x,y) &=& \bar c_{a}(x) \gamma^\nu c_a(x) ~ \bar q_{b}(y) \sigma_{\mu\nu} \gamma_5 q_b(y) \, ,
\\
\nonumber \theta^4_\mu(x,y) &=& \bar c_{a}(x) \sigma_{\mu\nu} \gamma_5 c_a(x) ~ \bar q_{b}(y) \gamma^\nu q_b(y) \, ,
\\
\nonumber \theta^5_\mu(x,y) &=& {\lambda^n_{ab}}{\lambda^n_{cd}} \times \left( \bar c_{a}(x) c_b(x) ~ \bar q_{c}(y) \gamma_\mu \gamma_5 q_d(y) \right) \, ,
\\
\nonumber \theta^6_\mu(x,y) &=& {\lambda^n_{ab}}{\lambda^n_{cd}} \times \left( \bar c_{a}(x) \gamma_\mu \gamma_5 c_b(x) ~ \bar q_{c}(y) q_d(y) \right) \, ,
\\
\nonumber \theta^7_\mu(x,y) &=& {\lambda^n_{ab}}{\lambda^n_{cd}} \times \left( \bar c_{a}(x) \gamma^\nu c_b(x) ~ \bar q_{c}(y) \sigma_{\mu\nu} \gamma_5 q_d(y) \right) \, ,
\\
\nonumber \theta^8_\mu(x,y) &=& {\lambda^n_{ab}}{\lambda^n_{cd}} \times \left( \bar c_{a}(x) \sigma_{\mu\nu} \gamma_5 c_b(x) ~ \bar q_{c}(y) \gamma^\nu q_d(y) \right) \, .
\end{eqnarray}
The former four $\theta^{1,2,3,4}_\mu$ have the color structure $[\bar c c]_{\mathbf{1}_c}[\bar q q]_{\mathbf{1}_c}\rightarrow[c \bar c q \bar q]_{\mathbf{1}_c}$, and the latter four $\theta^{5,6,7,8}_\mu$ have the color structure $[\bar c c]_{\mathbf{8}_c}[\bar q q]_{\mathbf{8}_c}\rightarrow[c \bar c q \bar q]_{\mathbf{1}_c}$.

\subsection{Fierz rearrangement}
\label{sec:fierz}

The Fierz rearrangement~\cite{fierz} of the Dirac and color indices has been systematically applied to study light baryon and tetraquark operators/currents in Refs.~\cite{Chen:2006hy,Chen:2006zh,Chen:2007xr,Chen:2013jra,Chen:2008qv,Chen:2009sf,Chen:2019wjd}. All the necessary equations can be found in Sec.~3.3.2 of Ref.~\cite{Chen:2016qju}. More studies can be found in Refs.~\cite{Nieves:2003in,Padmanath:2015era}. In the present study we apply it to relate the above three types of tetraquark currents.

The Fierz rearrangement is usually applied to local operators/currents. However, it is actually a matrix identity, and is valid if the same quark field in the initial and final operators is at the same location. As an example, we can apply the Fierz rearrangement to transform the non-local current with the quark fields $\eta(x^\prime,x;y^\prime,y) = [q(x^\prime) c(x)][\bar q(y^\prime) \bar c(y)]$ into a combination of several non-local currents with the quark fields at same locations $\theta(y,x;y^\prime,x^\prime) = [\bar c(y) c(x)][\bar q(y^\prime) q(x^\prime)]$. 

To apply it to study the decay process, we need to add two overall dynamical processes in the first and third steps:
\begin{eqnarray}
\eta(x,y) &=& ~\,[q(x)~c(x)]~\times~[\bar q(y)~\bar c(y)]
\\ \nonumber \rightarrow \eta(x^\prime,x;y^\prime,y)   &=& ~[q(x^\prime)~c(x)]~\times~[\bar q(y^\prime)~\bar c(y)]
\\ \nonumber \rightarrow \theta(y,x;y^\prime,x^\prime) &=& ~~[\bar c(y)~c(x)]~\times~[\bar q(y^\prime)~q(x^\prime)]
\\ \nonumber \rightarrow \theta(x^{\prime\prime};y^{\prime\prime}) &=& [\bar c(x^{\prime\prime})~c(x^{\prime\prime})] + [\bar q(y^{\prime\prime})~q(y^{\prime\prime})] \, ,
\end{eqnarray}
which will be detailedly discussed in the next section. The second step is the Fierz rearrangement whose explicit expressions are given as follows.

\begin{widetext}
Altogether, we obtain the following relation between the currents $\eta^i_\mu(x^\prime,x;y^\prime,y)$ and $\theta^i_\mu(y,x;y^\prime,x^\prime)$:
\begin{equation}
\left(\begin{array}{c}
\eta^1_\mu
\\
\eta^2_\mu
\\
\eta^3_\mu
\\
\eta^4_\mu
\\
\eta^5_\mu
\\
\eta^6_\mu
\\
\eta^7_\mu
\\
\eta^8_\mu
\end{array}\right)
=
\left(\begin{array}{cccccccc}
-{1/2} & {1/2} & -{{\rm i}/2} & {{\rm i}/2} & 0         & 0          & 0         & 0
\\
-{1/6} & {1/6} & -{{\rm i}/6} & {{\rm i}/6} & -{1/4} & {1/4} & -{{\rm i}/4} & {{\rm i}/4}
\\
-{3{\rm i}/2} & -{3{\rm i}/2} & {1/2} & {1/2} & {0} & {0} & {0} & {0}
\\
-{{\rm i}/2} & -{{\rm i}/2} & {1/6} & {1/6} & -{3{\rm i}/4} & -{3{\rm i}/4} & {1/4} & {1/4}
\\
-{1/2} & -{1/2} & {{\rm i}/2} & {{\rm i}/2} & {0} & {0} & {0} & {0}
\\
-{1/6} & -{1/6} & {{\rm i}/6} & {{\rm i}/6} & -{1/4} & -{1/4} & {{\rm i}/4} & {{\rm i}/4}
\\
-{3{\rm i}/2} & {3{\rm i}/2} & -{1/2} & {1/2} & {0} & {0} & {0} & {0}
\\
-{{\rm i}/2} & {{\rm i}/2} & -{1/6} & {1/6} & -{3{\rm i}/4} & {3{\rm i}/4} & -{1/4} & {1/4}
\end{array}\right)
\times
\left(\begin{array}{c}
\theta^1_\mu
\\
\theta^2_\mu
\\
\theta^3_\mu
\\
\theta^4_\mu
\\
\theta^5_\mu
\\
\theta^6_\mu
\\
\theta^7_\mu
\\
\theta^8_\mu
\end{array}\right) \, ,
\label{eq:fierz1}
\end{equation}
the following relation between $\eta^i_\mu(x^\prime,x;y^\prime,y)$ and $\xi^i_\mu(y,x^\prime;y^\prime,x)$:
\begin{equation}
\left(\begin{array}{c}
\eta^1_\mu
\\
\eta^2_\mu
\\
\eta^3_\mu
\\
\eta^4_\mu
\\
\eta^5_\mu
\\
\eta^6_\mu
\\
\eta^7_\mu
\\
\eta^8_\mu
\end{array}\right)
=
\left(\begin{array}{cccccccc}
{1/6} & 0 & 0 & -{{\rm i}/6} & {1/4} & 0 & 0 & -{{\rm i}/4}
\\
{1/2} & 0 & 0 & -{{\rm i}/2} & 0 & 0 & 0 & 0
\\
0 & -{1/6} & {{\rm i}/2} & 0 & 0 & -{1/4} & {3{\rm i}/4} & 0
\\
0 & -{1/2} & {3{\rm i}/2} & 0 & 0 & 0 & 0 & 0
\\
0 & {{\rm i}/6} & -{1/6} & 0 & 0 & {{\rm i}/4} & -{1/4} & 0
\\
0 & {{\rm i}/2} & -{1/2} & 0 & 0 & 0 & 0 & 0
\\
-{{\rm i}/2} & 0 & 0 & {1/6} & -{3{\rm i}/4} & 0 & 0 & {1/4}
\\
-{3{\rm i}/2} & 0 & 0 & {1/2} & 0 & 0 & 0 & 0
\end{array}\right)
\times
\left(\begin{array}{c}
\xi^1_\mu
\\
\xi^2_\mu
\\
\xi^3_\mu
\\
\xi^4_\mu
\\
\xi^5_\mu
\\
\xi^6_\mu
\\
\xi^7_\mu
\\
\xi^8_\mu
\end{array}\right) \, ,
\label{eq:fierz2}
\end{equation}
the following relation among $\eta^i_\mu(x^\prime,x;y^\prime,y)$, $\xi^{1,2,3,4}_\mu(y,x^\prime;y^\prime,x)$, and $\theta^{1,2,3,4}_\mu(y,x;y^\prime,x^\prime)$:
\begin{equation}
\left(\begin{array}{c}
\eta^1_\mu
\\
\eta^2_\mu
\\
\eta^3_\mu
\\
\eta^4_\mu
\\
\eta^5_\mu
\\
\eta^6_\mu
\\
\eta^7_\mu
\\
\eta^8_\mu
\end{array}\right)
=
\left(\begin{array}{cccccccc}
0 & 0 & 0 & 0 & -{1/2} & {1/2} & -{{\rm i}/2} & {{\rm i}/2}
\\
{1/2} & 0 & 0 & -{{\rm i}/2} & 0 & 0 & 0 & 0
\\
0 & 0 & 0 & 0 & -{3{\rm i}/2} & -{3{\rm i}/2} & {1/2} & {1/2}
\\
0 & -{1/2} & {3{\rm i}/2} & 0 & 0 & 0 & 0 & 0
\\
0 & 0 & 0 & 0 & -{1/2} & -{1/2} & {{\rm i}/2} & {{\rm i}/2}
\\
0 & {{\rm i}/2} & -{1/2} & 0 & 0 & 0 & 0 & 0
\\
0 & 0 & 0 & 0 & -{3{\rm i}/2} & {3{\rm i}/2} & -{1/2} & {1/2}
\\
-{3{\rm i}/2} & 0 & 0 & {1/2} & 0 & 0 & 0 & 0
\end{array}\right)
\times
\left(\begin{array}{c}
\xi^1_\mu
\\
\xi^2_\mu
\\
\xi^3_\mu
\\
\xi^4_\mu
\\
\theta^1_\mu
\\
\theta^2_\mu
\\
\theta^3_\mu
\\
\theta^4_\mu
\end{array}\right) \, ,
\label{eq:fierz3}
\end{equation}
and the following relation between $\xi^i_\mu(y,x^\prime;y^\prime,x)$ and $\theta^i_\mu(y,x;y^\prime,x^\prime)$:
\begin{equation}
\left(\begin{array}{c}
\xi^1_\mu
\\
\xi^2_\mu
\\
\xi^3_\mu
\\
\xi^4_\mu
\\
\xi^5_\mu
\\
\xi^6_\mu
\\
\xi^7_\mu
\\
\xi^8_\mu
\end{array}\right)
=
\left(\begin{array}{cccccccc}
{1/6} & -{1/6} & -{{\rm i}/6} & {{\rm i}/6} & {1/4} & -{1/4} & -{{\rm i}/4} & {{\rm i}/4}
\\
{{\rm i}/2} & {{\rm i}/2} & {1/6} & {1/6} & {3{\rm i}/4} & {3{\rm i}/4} & {1/4} & {1/4}
\\
-{1/6} & -{1/6} & -{{\rm i}/6} & -{{\rm i}/6} & -{1/4} & -{1/4} & -{{\rm i}/4} & -{{\rm i}/4}
\\
-{{\rm i}/2} & {{\rm i}/2} & {1/6} & -{1/6} & -{3{\rm i}/4} & {3{\rm i}/4} & {1/4} & -{1/4}
\\
{8/9} & -{8/9} & -{8{\rm i}/9} & {8{\rm i}/9} & -{1/6} & {1/6} & {{\rm i}/6} & -{{\rm i}/6}
\\
{8{\rm i}/3} & {8{\rm i}/3} & {8/9} & {8/9} & -{{\rm i}/2} & -{{\rm i}/2} & -{1/6} & -{1/6}
\\
-{8/9} & -{8/9} & -{8{\rm i}/9} & -{8{\rm i}/9} & {1/6} & {1/6} & {{\rm i}/6} & {{\rm i}/6}
\\
-{8{\rm i}/3} & {8{\rm i}/3} & {8/9} & -{8/9} & {{\rm i}/2} & -{{\rm i}/2} & -{1/6} & {1/6}
\end{array}\right)
\times
\left(\begin{array}{c}
\theta^1_\mu
\\
\theta^2_\mu
\\
\theta^3_\mu
\\
\theta^4_\mu
\\
\theta^5_\mu
\\
\theta^6_\mu
\\
\theta^7_\mu
\\
\theta^8_\mu
\end{array}\right) \, .
\label{eq:fierz4}
\end{equation}
\end{widetext}

\subsection{Isospin of the $X(3872)$ and decay constants}
\label{sec:current4}

In the present study we shall first use isoscalar tetraquark currents to study decay properties of the $X(3872)$, for example,
\begin{eqnarray}
\eta^1_\mu(I=0) &=& {1\over\sqrt2} \times \left( \eta^1_\mu([uc][\bar u \bar c]) + \eta^1_\mu([dc][\bar d \bar c]) \right)
\\ \nonumber &=& {1\over\sqrt2} \times \Big( u_a^{\rm T}\mathbb{C}\gamma_\mu c_b ~ \bar u_{a} \gamma_5\mathbb{C}\bar c_{b}^{\rm T} + \{ \gamma_\mu \leftrightarrow \gamma_5 \}
\\ \nonumber && ~~~~~~~~~~ +~ \{ u/\bar u \rightarrow d/\bar d \}  \Big) \, .
\end{eqnarray}
Hence, we need couplings of light isoscalar meson operators to light isoscalar meson states, which are summarized in Table~\ref{tab:coupling}. We also need couplings of charmonium operators to charmonium states as well as those of charmed meson operators to charmed meson states, which are also summarized in Table~\ref{tab:coupling}. We refer to Ref.~\cite{Chen:2019wjd} for detailed discussions.

Since light scalar mesons have a complicated nature~\cite{Pelaez:2015qba}, couplings of the light scalar-isoscalar meson operator $P^{S} = \left(\bar u u + \bar d d\right)/\sqrt2$ to $f_0$ mesons are quite ambiguous, where $f_0$ can be either the $f_0(500)$ or $f_0(1370)$, etc. In this paper we shall simply use the $f_0(500)$ meson to estimate relevant partial decay widths, whose coupling to $P^{S}$ is assumed to be
\begin{equation}
\langle 0 | {\bar u u + \bar d d \over \sqrt2} | f_0(p) \rangle = m_{f_0} f_{f_0} \, .
\end{equation}
In the present study we simply average among the decay constants $f_{\chi_{c0}}$ and $f_{D_0^{*}}$ to obtain
\begin{equation}
f_{f_0} \sim 380~{\rm MeV} \, .
\end{equation}

The isospin breaking effect of the $X(3872)$ is significant and important to understand its nature. There have been many studies on this, and we refer to reviews~\cite{Liu:2019zoy,Lebed:2016hpi,Esposito:2016noz,Guo:2017jvc,Ali:2017jda,Olsen:2017bmm,Karliner:2017qhf,Brambilla:2019esw} for detailed discussions. In the present study we shall study this effect by freely choosing the quark content of the $X(3872)$~\cite{Maiani:2004vq,Navarra:2006nd,Matheus:2006xi}, for example,
\begin{eqnarray}
\eta^1_\mu(\theta/\theta^\prime) &=& \cos\theta~\eta^1_\mu([uc][\bar u \bar c]) + \sin\theta~\eta^1_\mu([dc][\bar d \bar c])
\\ \nonumber &=& \cos\theta \times \left( u_a^{\rm T}\mathbb{C}\gamma_\mu c_b ~ \bar u_{a} \gamma_5\mathbb{C}\bar c_{b}^{\rm T} + \{ \gamma_\mu \leftrightarrow \gamma_5 \} \right)
\\ \nonumber && +~\sin\theta \times \{ u/\bar u \rightarrow d/\bar d \}
\\ \nonumber &\Rightarrow& \cos\theta^\prime~\eta^1_\mu(I=0) + \sin\theta^\prime~\eta^1_\mu(I=1) \, ,
\end{eqnarray}
where $\theta$ and $\theta^\prime$ are the two related mixing angles. We shall fine-tune them to be different from $\theta = 45^\circ/\theta^\prime = 0^\circ$ in Sec.~\ref{sec:isospin}, so that the $X(3872)$ is assumed not to be a purely isoscalar state. To study this, we need couplings of light isovector meson operators to light isovector meson states, which are also summarized in Table~\ref{tab:coupling}.

\begin{table*}[hbt]
\begin{center}
\renewcommand{\arraystretch}{1.5}
\caption{Couplings of meson operators to meson states. All the light isovector meson operators $J^{S/P/V/A/T}_{(\mu\nu)}$ have the quark content $\bar q \Gamma q = \left(\bar u \Gamma u - \bar d \Gamma d\right)/\sqrt2$, and all the light isoscalar meson operators $P^{S/P/V/A/T}_{(\mu\nu)}$ have the quark content $\bar q \Gamma q = \left(\bar u \Gamma u + \bar d \Gamma d\right)/\sqrt2$. Color indices are omitted for simplicity.}
\begin{tabular}{ c | c | c | c | c | c}
\hline\hline
~~~Operators~~~ & ~~$I^GJ^{PC}$~~ & ~~~~~~~~Mesons~~~~~~~~ & ~~$I^GJ^{PC}$~~ & ~~~Couplings~~~ & ~~~~~~~~Decay Constants~~~~~~~~
\\ \hline\hline
$J^{S} = \bar q q$ & $1^-0^{++}$ & -- & $1^-0^{++}$ & -- & --
\\ \hline
$J^{P} = \bar q {\rm i}\gamma_5 q$ & $1^-0^{-+}$ &  $\pi^0$  & $1^-0^{-+}$ &  $\langle 0 | J^{P} | \pi^0 \rangle = \lambda_\pi$  &  $\lambda_\pi = {f_\pi m_\pi^2 \over m_u + m_d}$
\\ \hline
$J^{V}_\mu = \bar q \gamma_\mu q$ & $1^+1^{--}$ &  $\rho^0$  & $1^+1^{--}$ &  $\langle 0 | J^{V}_\mu | \rho^0 \rangle = m_\rho f_{\rho} \epsilon_\mu$  &  $f_{\rho} = 216$~MeV~\mbox{\cite{Jansen:2009hr}}
\\ \hline
\multirow{2}{*}{$J^{A}_\mu = \bar q \gamma_\mu \gamma_5 q$} & \multirow{2}{*}{$1^-1^{++}$} & $\pi^0$  & $1^-0^{-+}$ & $\langle 0 | J^{A}_\mu | \pi^0 \rangle = {\rm i} p_\mu f_{\pi}$  &  $f_{\pi} = 130.2$~MeV~\mbox{\cite{pdg}}
\\ \cline{3-6}
                                                         & &  $a_1(1260)$  & $1^-1^{++}$ & $\langle 0 | J^{A}_\mu | a_1 \rangle = m_{a_1} f_{a_1} \epsilon_\mu $  &  $f_{a_1} = 254$~MeV~\mbox{\cite{Wingate:1995hy}}
\\ \hline
\multirow{2}{*}{$J^{T}_{\mu\nu} = \bar q \sigma_{\mu\nu} q$} & \multirow{2}{*}{$1^+1^{\pm-}$} &  $\rho^0$ & $1^+1^{--}$   &  $\langle 0 | J^{T}_{\mu\nu} | \rho^0 \rangle = {\rm i} f^T_{\rho} (p_\mu\epsilon_\nu - p_\nu\epsilon_\mu) $  &  $f_{\rho}^T = 159$~MeV~\mbox{\cite{Jansen:2009hr}}
\\ \cline{3-6}
                                                             & &  $b_1(1235)$  & $1^+1^{+-}$ &  $\langle 0 | J^{T}_{\mu\nu} | b_1 \rangle = {\rm i} f^T_{b_1} \epsilon_{\mu\nu\alpha\beta} \epsilon^\alpha p^\beta $  &  $f_{b_1}^T = 180$~MeV~\mbox{\cite{Ball:1996tb}}
\\ \hline\hline
$P^{S} = \bar q q$                & $0^+0^{++}$                  & $f_0(500)$~(?) & $0^+0^{++}$ & $\langle 0 | P^S | f_0 \rangle = m_{f_0} f_{f_0}$          & $f_{f_0} \sim 380$~MeV~(?)
\\ \hline
$P^{P} = \bar q {\rm i}\gamma_5 q$      & $0^+0^{-+}$                  & $\eta$               & $0^+0^{-+}$ & -- & -- 
\\ \hline
$P^{V}_\mu = \bar q \gamma_\mu q$ & $0^-1^{--}$                  & $\omega$             & $0^-1^{--}$ & $\langle 0 | P^{V}_\mu | \omega \rangle = m_\omega f_\omega \epsilon_\mu$
                               & $f_\omega \approx f_{\rho} = 216$~MeV~\mbox{\cite{Jansen:2009hr}}
\\ \hline
\multirow{2}{*}{$P^{A}_\mu = \bar q \gamma_\mu \gamma_5 q$}
                               & \multirow{2}{*}{$0^+1^{++}$} & $\eta$               & $0^+0^{-+}$ & $\langle 0 | P^{A}_\mu | \eta \rangle = {\rm i} p_\mu f_{\eta}$                    & $f_{\eta} = 97$~MeV~\mbox{\cite{Ottnad:2017bjt,Guo:2015xva}}
\\ \cline{3-6}
                               &                              & $f_1(1285)$          & $0^+1^{++}$ & -- & -- 
\\ \hline
\multirow{2}{*}{$P^{T}_{\mu\nu} = \bar q \sigma_{\mu\nu} q$}
                               & \multirow{2}{*}{$0^-1^{\pm-}$} & $\omega$             & $0^-1^{--}$ & $\langle 0 | P^{T}_{\mu\nu} | \omega \rangle = {\rm i} f^T_{\omega} (p_\mu\epsilon_\nu - p_\nu\epsilon_\mu) $
                               &  $f^T_{\omega} \approx f_{\rho}^T = 159$~MeV~\mbox{\cite{Jansen:2009hr}}
\\ \cline{3-6}
                               &                              & $h_1(1170)$          & $0^-1^{+-}$ & $\langle 0 | P^{T}_{\mu\nu} | h_1 \rangle = {\rm i} f^T_{h_1} \epsilon_{\mu\nu\alpha\beta} \epsilon^\alpha p^\beta $
                               &  $f_{h_1}^T \approx f_{b_1}^T = 180$~MeV~\mbox{\cite{Ball:1996tb}}
\\ \hline\hline
$I^{S} = \bar c c$                & $0^+0^{++}$                  & $\chi_{c0}(1P)$      & $0^+0^{++}$ & $\langle 0 | I^S | \chi_{c0} \rangle = m_{\chi_{c0}} f_{\chi_{c0}}$          & $f_{\chi_{c0}} = 343$~MeV~\mbox{\cite{Veliev:2010gb}}
\\ \hline
$I^{P} = \bar c {\rm i}\gamma_5 c$      & $0^+0^{-+}$                  & $\eta_c$             & $0^+0^{-+}$ & $\langle 0 | I^{P} | \eta_c \rangle = \lambda_{\eta_c}$                      & $\lambda_{\eta_c} = {f_{\eta_c} m_{\eta_c}^2 \over 2 m_c}$
\\ \hline
$I^{V}_\mu = \bar c \gamma_\mu c$ & $0^-1^{--}$                  & $J/\psi$             & $0^-1^{--}$ & $\langle0| I^{V}_\mu | J/\psi \rangle = m_{J/\psi} f_{J/\psi} \epsilon_\mu$  & $f_{J/\psi} = 418$~MeV~\mbox{\cite{Becirevic:2013bsa}}
\\ \hline
\multirow{2}{*}{$I^{A}_\mu = \bar c \gamma_\mu \gamma_5 c$}
                               & \multirow{2}{*}{$0^+1^{++}$}    & $\eta_c$             & $0^+0^{-+}$ & $\langle 0 | I^{A}_\mu | \eta_c \rangle = {\rm i} p_\mu f_{\eta_c}$                & $f_{\eta_c} = 387$~MeV~\mbox{\cite{Becirevic:2013bsa}}
\\ \cline{3-6}
                               &                              & $\chi_{c1}(1P)$      & $0^+1^{++}$ & $\langle 0 | I^{A}_\mu | \chi_{c1} \rangle = m_{\chi_{c1}} f_{\chi_{c1}} \epsilon_\mu $
                               &  $f_{\chi_{c1}} = 335$~MeV~\mbox{\cite{Novikov:1977dq}}
\\ \hline
\multirow{2}{*}{$I^{T}_{\mu\nu} = \bar c \sigma_{\mu\nu} c$}
                               & \multirow{2}{*}{$0^-1^{\pm-}$}  & $J/\psi$             & $0^-1^{--}$ & $\langle 0 | I^{T}_{\mu\nu} | J/\psi \rangle = {\rm i} f^T_{J/\psi} (p_\mu\epsilon_\nu - p_\nu\epsilon_\mu) $
                               &  $f_{J/\psi}^T = 410$~MeV~\mbox{\cite{Becirevic:2013bsa}}
\\ \cline{3-6}
                               &                              & $h_c(1P)$            & $0^-1^{+-}$ & $\langle 0 | I^{T}_{\mu\nu} | h_c \rangle = {\rm i} f^T_{h_c} \epsilon_{\mu\nu\alpha\beta} \epsilon^\alpha p^\beta $
                               &  $f_{h_c}^T = 235$~MeV~\mbox{\cite{Becirevic:2013bsa}}
\\ \hline\hline
$O^{S} = \bar q c$                & $0^{+}$                   & $D_0^{*}$           & $0^{+}$  & $\langle 0 | O^{S} | D_0^{*} \rangle = m_{D_0^{*}} f_{D_0^{*}}$             &  $f_{D_0^{*}} = 410$~MeV~\mbox{\cite{Narison:2015nxh}}
\\ \hline
$O^{P} = \bar q {\rm i}\gamma_5 c$      & $0^{-}$                   & $D$                & $0^{-}$  & $\langle 0 | O^{P} | D \rangle = \lambda_D$                                &  $\lambda_D = {f_D m_D^2 \over {m_c + m_d}}$
\\ \hline
$O^{V}_\mu = \bar c \gamma_\mu q$ & $1^{-}$                   & $\bar D^{*}$        & $1^{-}$  & $\langle0| O^{V}_\mu | \bar D^{*} \rangle = m_{D^*} f_{D^*} \epsilon_\mu$   &  $f_{D^*} = 253$~MeV~\mbox{\cite{Chang:2018aut}}
\\ \hline
\multirow{2}{*}{$O^{A}_\mu = \bar c \gamma_\mu \gamma_5 q$}
                               & \multirow{2}{*}{$1^{+}$}      & $\bar D$         & $0^{-}$  & $\langle 0 | O^{A}_\mu | \bar D \rangle = {\rm i} p_\mu f_{D}$                 &  $f_{D} = 211.9$~MeV~\mbox{\cite{pdg}}
\\ \cline{3-6}
                                  &                            & $D_1$                & $1^{+}$  & $\langle 0 | O^{A}_\mu | D_1 \rangle = m_{D_1} f_{D_1} \epsilon_\mu $        &  $f_{D_1} = 356$~MeV~\mbox{\cite{Narison:2015nxh}}
\\ \hline
\multirow{2}{*}{$O^{T}_{\mu\nu} = \bar q \sigma_{\mu\nu} c$}
                               & \multirow{2}{*}{$1^{\pm}$}    & $\bar D^{*}$        & $1^{-}$  &  $\langle 0 | O^{T}_{\mu\nu} | D^{*} \rangle = {\rm i} f_{D^*}^T (p_\mu\epsilon_\nu - p_\nu\epsilon_\mu) $
                               &  $f_{D^*}^T \approx 220$~MeV~\mbox{\cite{Chen:2019wjd}}
\\ \cline{3-6}
                               &                               &  --                  & $1^{+}$  &  --  &  --
\\ \hline\hline
\end{tabular}
\label{tab:coupling}
\end{center}
\end{table*}

\section{Decay properties of the $X(3872)$ as a diquark-antidiquark state}
\label{sec:decaydiquark}

In this section and the next we shall use Eqs.~(\ref{eq:fierz1}-\ref{eq:fierz4}) derived in the previous section to study decay properties of the $X(3872)$ as a purely isoscalar state. Its two possible interpretations are: a) the compact tetraquark state of $J^{PC} = 1^{++}$ composed of a $J^P= 0^+$ diquark/antidiquark and a $J^P= 1^+$ antidiquark/diquark~\cite{Maiani:2004vq,Maiani:2014aja,Hogaasen:2005jv,Ebert:2005nc,Barnea:2006sd,Chiu:2006hd}, {\it i.e.}, $|0_{qc}1_{\bar q \bar c} ; 1^{++} \rangle$ defined in Eq.~(\ref{def:diquark}); and b) the $D \bar D^*$ hadronic molecular state of $J^{PC} = 1^{++}$~\cite{Voloshin:1976ap,Tornqvist:2004qy,Close:2003sg,Voloshin:2003nt,Wong:2003xk,Braaten:2003he,Swanson:2003tb}, {\it i.e.}, $| D \bar D^*; 1^{++} \rangle$ defined in Eq.~(\ref{def:molecule}).

In this section we investigate the former compact tetraquark interpretation using the isoscalar current $\eta^{\mathcal X}_\mu(x,y; I=0)$ defined in Eq.~(\ref{current:diquark}). We can transform it to $\theta_\mu^i(x,y; I=0)$ and $\xi_\mu^i(x,y; I=0)$ according to Eqs.~(\ref{eq:fierz1}-\ref{eq:fierz3}), through which we study decay properties of the $X(3872)$ as an isoscalar compact tetraquark state in the following subsections.

\subsection{$\eta^{\mathcal X}_\mu\big([qc][\bar q \bar c]\big) \rightarrow \theta_\mu^i \big([\bar c c] + [\bar q q]\big)$}
\label{sec:diquark1}

%
\begin{figure*}[hbt]
\begin{center}
\includegraphics[width=0.8\textwidth]{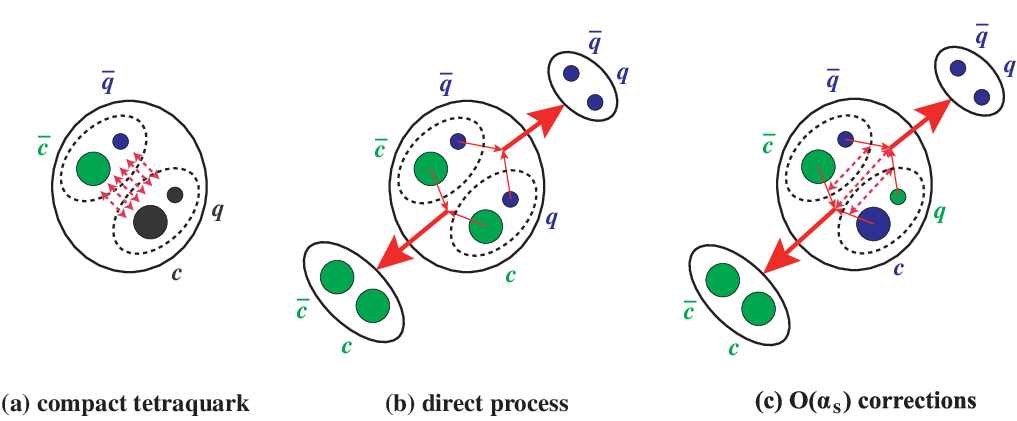}
\caption{The decay of a compact tetraquark state into one charmonium meson and one light meson, which can happen through either (b) a direct fall-apart process, or (c) a process with gluons exchanged.}
\label{fig:a}
\end{center}
\end{figure*}
%

As depicted in Fig.~\ref{fig:a}, when the $q$ and $\bar q$ quarks meet each other and the $c$ and $\bar c$ quarks meet each other at the same time, a compact tetraquark state decays into one charmonium meson and one light meson:
\begin{eqnarray}
&& [q(x) c(x)]~[\bar q(y) \bar c(y)]
\label{eq:change1}
\\ \nonumber &\Longrightarrow& [q(x \to y^\prime)~c(x \to x^\prime)]~[\bar q(y \to y^\prime)~\bar c(y \to x^\prime)]
\\ \nonumber &\Longrightarrow& [\bar c(x^\prime) c(x^\prime)] + [\bar q(y^\prime) q(y^\prime)] \, ,
\end{eqnarray}
The first process is a dynamical process, and the second process can be described through the transformation~(\ref{eq:fierz1}):
\begin{eqnarray}
&& \eta^{\mathcal X}_\mu(x,y;I=0)
\label{tran:diquark1}
\\ \nonumber &\Longrightarrow& - {1\over3}~\theta_\mu^1(x^\prime,y^\prime;I=0) + {1\over3}~\theta_\mu^2(x^\prime,y^\prime;I=0)
\\ \nonumber &&  - {{\rm i}\over3}~\theta_\mu^3(x^\prime,y^\prime;I=0) + {{\rm i}\over3}~\theta_\mu^4(x^\prime,y^\prime;I=0) + \cdots
\\ \nonumber &=& - {1\over3}~I^{S}(x^\prime)~P^{A}_\mu(y^\prime) + {1\over3}~I^{A}_\mu(x^\prime)~P^{S}(y^\prime)
\\ \nonumber &&  + {1\over6}~\epsilon_{\mu\nu\rho\sigma}~I^{V,\nu}(x^\prime)~P^{T,\rho\sigma}(y^\prime)
\\ \nonumber &&  - {1\over6}~\epsilon_{\mu\nu\rho\sigma}~I^{T,\rho\sigma}(x^\prime)~P^{V,\nu}(y^\prime)~+~\cdots \, ,
\end{eqnarray}
where we have used
\begin{equation}
\sigma_{\mu\nu} \gamma_5 = {{\rm i}\over2} \epsilon_{\mu\nu\rho\sigma} \sigma^{\rho\sigma} \, .
\end{equation}
In the above expression we keep only the direct fall-apart process described by $\theta_\mu^{1,2,3,4}$, but neglect the $\mathcal{O}(\alpha_s)$ corrections described by $\theta_\mu^{5,6,7,8}$.

Together with Table~\ref{tab:coupling}, we extract the following decay channels:
\begin{enumerate}

\item The decay of $|0_{qc}1_{\bar q \bar c} ; 1^{++} \rangle$ into $\chi_{c0} \eta$ is contributed by $I^{S} \times P^{A}_\mu$:
\begin{eqnarray}
&& \langle X(p,\epsilon) | \chi_{c0}(p_1)~\eta(p_2) \rangle
\\ \nonumber &\approx& - {{\rm i}c_1\over3}~m_{\chi_{c0}} f_{\chi_{c0}} f_{\eta}~\epsilon \cdot p_2 \equiv g_{\chi_{c0} \eta}~\epsilon \cdot p_2 \, ,
\end{eqnarray}
where $c_1$ is an overall factor, related to the coupling of $\eta^{\mathcal X}_\mu(x,y)$ to the $X(3872)$ as well as the dynamical process $(x,y) \Longrightarrow (x^\prime, y^\prime)$ shown in Fig.~\ref{fig:a}.
This decay is kinematically forbidden.

\item According to $I^{S} \times P^{A}_\mu$, $|0_{qc}1_{\bar q \bar c} ; 1^{++} \rangle$ can also decay into $\chi_{c0} f_1(1285)$:
\begin{equation}
|0_{qc}1_{\bar q \bar c} ; 1^{++} \rangle \rightarrow \chi_{c0} f_1 \, .
\end{equation}
This decay is kinematically forbidden.

\item Decays of $|0_{qc}1_{\bar q \bar c} ; 1^{++} \rangle$ into $\eta_c f_0(500)$ and $\chi_{c1} f_0(500)$ are both contributed by $I^{A}_\mu \times P^{S}$:
\begin{eqnarray}
&& \langle X(p,\epsilon) | \eta_c(p_1)~f_0(p_2) \rangle
\\ \nonumber &\approx& + {{\rm i}c_1\over3}~m_{f_0} f_{f_0} f_{\eta_c}~\epsilon \cdot p_1 \equiv g_{\eta_c f_0}~\epsilon \cdot p_1 \, ,
\\ && \langle X(p,\epsilon) | \chi_{c1}(p_1,\epsilon_1)~f_0(p_2) \rangle
\\ \nonumber &\approx& + {c_1\over3}~m_{f_0} f_{f_0} m_{\chi_{c1}} f_{\chi_{c1}}~\epsilon \cdot \epsilon_1 \equiv g_{\chi_{c1} f_0}~\epsilon \cdot \epsilon_1 \, .
\end{eqnarray}
Because it is difficult to observe the $f_0(500)$ in experiments, in the present study we shall calculate widths of the $|0_{qc}1_{\bar q \bar c} ; 1^{++} \rangle \rightarrow \eta_c f_0 \rightarrow \eta_c \pi \pi$ and $|0_{qc}1_{\bar q \bar c} ; 1^{++} \rangle \rightarrow \chi_{c1} f_0 \rightarrow \chi_{c1} \pi \pi$ processes.

\item The decay of $|0_{qc}1_{\bar q \bar c} ; 1^{++} \rangle$ into $J/\psi \omega$ is contributed by both $I^{V,\nu} \times P^{T,\rho\sigma}$ and $I^{T,\rho\sigma} \times P^{V,\nu}$:
\begin{eqnarray}
&& \langle X(p,\epsilon) | J/\psi(p_1,\epsilon_1)~\omega(p_2,\epsilon_2) \rangle
\\ \nonumber &\approx& -{{\rm i}c_1\over3}~\epsilon_{\mu\nu\rho\sigma} \epsilon^\mu \epsilon_1^\nu \epsilon_2^\rho p_2^\sigma~m_{J/\psi}f_{J/\psi}f_{\omega}^T
\\ \nonumber &&        -{{\rm i}c_1\over3}~\epsilon_{\mu\nu\rho\sigma} \epsilon^\mu \epsilon_1^\nu \epsilon_2^\rho  p_1^\sigma~m_{\omega}f_{\omega}f_{J/\psi}^T
\\ \nonumber &\equiv& g^A_{\psi \omega}~\epsilon_{\mu\nu\rho\sigma} \epsilon^\mu \epsilon_1^\nu \epsilon_2^\rho p_2^\sigma + g^B_{\psi \omega}~\epsilon_{\mu\nu\rho\sigma} \epsilon^\mu \epsilon_1^\nu \epsilon_2^\rho p_1^\sigma \, .
\end{eqnarray}
This decay is kinematically forbidden, but the $|0_{qc}1_{\bar q \bar c} ; 1^{++} \rangle \rightarrow J/\psi \omega \rightarrow J/\psi \pi\pi\pi$ process is kinematically allowed.

\item According to $I^{V,\nu} \times P^{T,\rho\sigma}$, $|0_{qc}1_{\bar q \bar c} ; 1^{++} \rangle$ can also decay into $J/\psi h_1(1170)$:
\begin{eqnarray}
&& \langle X(p,\epsilon) | J/\psi(p_1,\epsilon_1)~h_1(p_2,\epsilon_2) \rangle
\\ \nonumber &\approx& +{{\rm i}c_1\over6}~\epsilon_{\mu\nu\rho\sigma} \epsilon_{\rho\sigma\alpha\beta} \epsilon^\mu \epsilon_1^\nu \epsilon_2^\alpha  p_2^\beta~m_{J/\psi}f_{J/\psi}f_{h_1}^T
\\ \nonumber &\equiv& g_{\psi h_1}~\epsilon_{\mu\nu\rho\sigma} \epsilon_{\rho\sigma\alpha\beta} \epsilon^\mu \epsilon_1^\nu \epsilon_2^\alpha  p_2^\beta \, .
\end{eqnarray}
This decay is kinematically forbidden, but the $|0_{qc}1_{\bar q \bar c} ; 1^{++} \rangle \rightarrow J/\psi h_1 \rightarrow J/\psi \rho \pi \rightarrow J/\psi \pi\pi\pi$ process is kinematically allowed.

\item According to $I^{T,\rho\sigma} \times P^{V,\nu}$, $|0_{qc}1_{\bar q \bar c} ; 1^{++} \rangle$ can also decay into $h_c \omega$:
\begin{equation}
|0_{qc}1_{\bar q \bar c} ; 1^{++} \rangle \rightarrow h_c \omega \, .
\end{equation}
This decay is kinematically forbidden.

\end{enumerate}

Summarizing the above results, we obtain numerically
\begin{eqnarray}
\nonumber g_{\eta_c f_0} &\sim& + {\rm i} c_1~2.51 \times 10^{7}~{\rm MeV}^3 \, ,
\\ \nonumber g_{\chi_{c1} f_0} &\sim& + c_1~0.76 \times 10^{11}~{\rm MeV}^4 \, ,
\\ g^A_{\psi \omega} &=& - {\rm i} c_1~6.86 \times 10^{7}~{\rm MeV}^3 \, ,
\\ \nonumber g^B_{\psi \omega} &=& - {\rm i} c_1~2.31 \times 10^{7}~{\rm MeV}^3 \, ,
\\ \nonumber g_{\psi h_1} &=& + {\rm i} c_1~3.88 \times 10^{7}~{\rm MeV}^3 \, ,
\end{eqnarray}
from which we further obtain
\begin{eqnarray}
{\mathcal{B}(|0_{qc}1_{\bar q \bar c} ; 1^{++} \rangle \rightarrow \eta_c f_0 \rightarrow \eta_c \pi \pi)
\over
\mathcal{B}(|0_{qc}1_{\bar q \bar c} ; 1^{++} \rangle \rightarrow J/\psi \omega \rightarrow J/\psi \pi \pi \pi)} &\sim& 0.091 \, ,
\label{eq:diquarkbr1}
\\ \nonumber {\mathcal{B}(|0_{qc}1_{\bar q \bar c} ; 1^{++} \rangle \rightarrow \chi_{c1} f_0 \rightarrow \chi_{c1} \pi \pi)
\over
\mathcal{B}(|0_{qc}1_{\bar q \bar c} ; 1^{++} \rangle \rightarrow J/\psi \omega \rightarrow J/\psi \pi \pi \pi)} &\sim& 0.086 \, ,
\\ \nonumber {\mathcal{B}(|0_{qc}1_{\bar q \bar c} ; 1^{++} \rangle \rightarrow J/\psi h_1 \rightarrow J/\psi \pi \pi \pi)
\over
\mathcal{B}(|0_{qc}1_{\bar q \bar c} ; 1^{++} \rangle \rightarrow J/\psi \omega \rightarrow J/\psi \pi \pi \pi)} &=& 1.4 \times 10^{-3} \, .
\end{eqnarray}
Detailed calculations can be found in Appendix~\ref{sec:width}. 

\subsection{$\eta^{\mathcal X}_\mu\big([qc][\bar q \bar c]\big) \rightarrow \xi_\mu^i \big([\bar c q] + [\bar q c]\big)$}
\label{sec:diquark2}

%
\begin{figure*}[hbt]
\begin{center}
\includegraphics[width=0.8\textwidth]{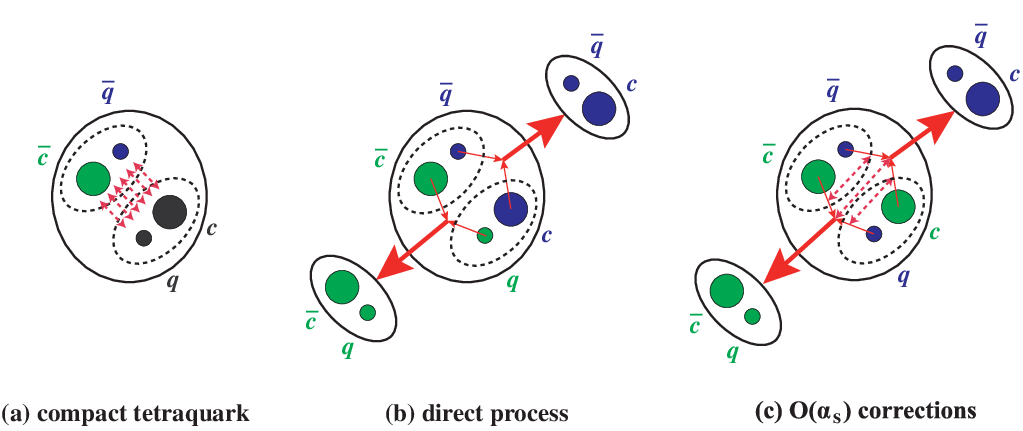}
\caption{The decay of a compact tetraquark state into two charmed mesons, which can happen through either (b) a direct fall-apart process, or (c) a process with gluons exchanged.}
\label{fig:b}
\end{center}
\end{figure*}
%

As depicted in Fig.~\ref{fig:b}, when the $c$ and $\bar q$ quarks meet each other and the $q$ and $\bar c$ quarks meet each other at the same time, a compact tetraquark state decays into two charmed mesons. This process can be described by the transformation~(\ref{eq:fierz2}):
\begin{eqnarray}
&& \eta^{\mathcal X}_\mu(x,y; I = 0)
\label{tran:diquark2}
\\ \nonumber &\Longrightarrow& - {1\over3}~\xi_\mu^1(x^\prime,y^\prime; I = 0) + {{\rm i}\over3}~\xi_\mu^4(x^\prime,y^\prime; I = 0) + \cdots
\\ \nonumber &=& + {{\rm i}\over3}~O^{V}_\mu(x^\prime)~O^{P}(y^\prime) + {{\rm i}\over3}~O^{A,\nu}(x^\prime)~O^{T}_{\mu\nu}(y^\prime)
\\ \nonumber &&  +~ c.c. ~+~ \cdots \, .
\end{eqnarray}
Again, we keep only the direct fall-apart process described by $\xi_\mu^{1,4}$, but neglect the $\mathcal{O}(\alpha_s)$ corrections described by $\xi_\mu^{5,8}$.

The decay of $|0_{qc}1_{\bar q \bar c} ; 1^{++} \rangle$ into the $D \bar D^{*}$ final state is contributed by both $O^{V}_\mu \times O^{P}$ and $O^{A,\nu} \times O^{T}_{\mu\nu}$:
\begin{eqnarray}
&& \langle X(p,\epsilon) | D^0(p_1) \bar D^{*0}(p_2,\epsilon_2) \rangle
\\ \nonumber &\approx& + {{\rm i}c_2\over3\sqrt2}~\lambda_D m_{D^*} f_{D^*}~\epsilon \cdot \epsilon_2
\\ \nonumber && ~~~~~  - {{\rm i}c_2\over3\sqrt2}~f_{D} f^T_{D^*}~(\epsilon \cdot p_2 ~ \epsilon_2 \cdot p_1 - p_1 \cdot p_2~\epsilon \cdot \epsilon_2)
\\ \nonumber &\equiv& g^S_{D \bar D^{*}}~\epsilon \cdot \epsilon_2 + g^D_{D \bar D^{*}}~(\epsilon \cdot p_2 ~ \epsilon_2 \cdot p_1 - p_1 \cdot p_2~\epsilon \cdot \epsilon_2) \, ,
\end{eqnarray}
where $c_2$ is an overall factor, related to the coupling of $\eta^{\mathcal X}_\mu(x,y)$ to the $X(3872)$ as well as the dynamical process $(x,y) \Longrightarrow (x^\prime, y^\prime)$ shown in Fig.~\ref{fig:b}. This decay might be kinematically forbidden, but the $|0_{qc}1_{\bar q \bar c} ; 1^{++} \rangle \rightarrow D^0 \bar D^{*0} + D^{*0} \bar D^0 \rightarrow D^0 \bar D^0 \pi^0$ process is surely kinematically allowed. The two coupling constants $g^S_{D \bar D^{*}}$ and $g^D_{D \bar D^{*}}$ are defined for the $S$- and $D$-wave $X(3872) \to D \bar D^{*}$ decays:
\begin{eqnarray}
\mathcal{L}^S_{D \bar D^{*}} &=& g^S_{D \bar D^{*}}~X^{\mu}~D^0~\bar D^{*0}_{\mu} + \cdots \, ,
\label{lag:DDsS}
\\ \mathcal{L}^D_{D \bar D^{*}} &=& g^D_{D \bar D^{*}} \times \left( g^{\mu\sigma}g^{\nu\rho} - g^{\mu\nu}g^{\rho\sigma} \right)
\label{lag:DDsD}
\\ \nonumber && ~~~~~ \times X^{\mu}~\partial_\rho D^0~\partial_\sigma \bar D^{*0}_{\nu} + \cdots \, .
\end{eqnarray}

Numerically, we obtain
\begin{eqnarray}
g^S_{D \bar D^{*}} &=& + {\rm i}c_2~0.69 \times 10^{11}~{\rm MeV}^4 \, ,
\\ \nonumber g^D_{D \bar D^{*}} &=& - {\rm i}c_2~1.10 \times 10^{4}~{\rm MeV}^2 \, .
\end{eqnarray}
Comparing this decay with the $|0_{qc}1_{\bar q \bar c} ; 1^{++} \rangle \rightarrow J/\psi \omega \rightarrow J/\psi \pi \pi \pi$ decay studied in the previous subsection, we further obtain
\begin{eqnarray}
\nonumber && {\mathcal{B}(|0_{qc}1_{\bar q \bar c} ; 1^{++} \rangle \rightarrow D^0 \bar D^{*0} + D^{*0} \bar D^0 \rightarrow D^0 \bar D^0 \pi^0)
\over
\mathcal{B}(|0_{qc}1_{\bar q \bar c} ; 1^{++} \rangle \rightarrow J/\psi \omega \rightarrow J/\psi \pi \pi \pi)}
\\ &=& 0.32 \times {c_2^2 \over c_1^2} \, .
\end{eqnarray}
Detailed calculations can also be found in Appendix~\ref{sec:width}. As proposed in Ref.~\cite{Maiani:2017kyi}, when the $X(3872)$ decays, a constituent of the diquark must tunnel through the barrier of the diquark-antidiquark potential. However, this tunnelling for heavy quarks is exponentially suppressed compared to that for light quarks, so the compact tetraquark couplings are expected to favour the open charm modes with respect to charmonium ones. Accordingly, $c_2$ may be significantly larger than $c_1$, so that $|0_{qc}1_{\bar q \bar c} ; 1^{++} \rangle$ may mainly decay into the $D^0 \bar D^0 \pi^0$ final state.

\subsection{$\eta^{\mathcal X}_\mu\big([qc][\bar q \bar c]\big) \rightarrow \theta_\mu^{1,2,3,4}\big([\bar c c] + [\bar q q]\big) + \xi_\mu^{1,2,3,4}\big([\bar c q] + [\bar q c]\big)$}
\label{sec:diquark3}

If the above two processes investigated in Sec.~\ref{sec:diquark1} and Sec.~\ref{sec:diquark2} happen at the same time, {\it i.e.}, $|0_{qc}1_{\bar q \bar c} ; 1^{++} \rangle$ decays into one charmonium meson and one light meson as well as two charmed mesons simultaneously, we can use the transformation~(\ref{eq:fierz3}), which contains the color-singlet-color-singlet currents $\theta_\mu^{1,2,3,4}$ and $\xi_\mu^{1,2,3,4}$ together:
\begin{eqnarray}
&& \eta^{\mathcal X}_\mu(x,y; I=0)
\label{tran:diquark3}
\\ \nonumber &\Longrightarrow& - {1\over2}~\theta_\mu^1(x^\prime,y^\prime; I=0)              + {1\over2}~\theta_\mu^2(x^\prime,y^\prime; I=0)
\\ \nonumber &&                - {{\rm i}\over2}~\theta_\mu^3(x^\prime,y^\prime; I=0)              + {{\rm i}\over2}~\theta_\mu^4(x^\prime,y^\prime; I=0)
\\ \nonumber &&                - {1\over2}~\xi_\mu^1(x^{\prime\prime},y^{\prime\prime}; I=0) + {{\rm i}\over2}~\xi_\mu^4(x^{\prime\prime},y^{\prime\prime}; I=0) \, .
\end{eqnarray}
In the above expression we keep all terms, and there is no $\cdots$ any more. Comparing this equation with Eqs.~(\ref{tran:diquark1}) and (\ref{tran:diquark2}), we obtain the same relative branching ratios as Sec.~\ref{sec:diquark1} and Sec.~\ref{sec:diquark2}, just with the overall factors $c_1$ and $c_2$ replaced by others.

\section{Decay properties of the $X(3872)$ as a hadronic molecular state}
\label{sec:decaymolecule}

Another possible interpretation of the $X(3872)$ is the $D \bar D^*$ hadronic molecular state of $J^{PC} = 1^{++}$~\cite{Voloshin:1976ap,Tornqvist:2004qy,Close:2003sg,Voloshin:2003nt,Wong:2003xk,Braaten:2003he,Swanson:2003tb}, {\it i.e.}, $| D \bar D^*; 1^{++} \rangle$ defined in Eq.~(\ref{def:molecule}). Its relevant isoscalar current $\xi^{\mathcal X}_\mu(x,y; I=0)$ is given in Eq.~(\ref{current:molecule}). We can transform it to $\theta_\mu^i(x,y; I=0)$ according to Eq.~(\ref{eq:fierz4}), through which we study decay properties of the $X(3872)$ as an isoscalar hadronic molecular state in the following subsections.

\subsection{$\xi^{\mathcal X}_\mu\big([\bar c q][\bar q c]\big) \longrightarrow \theta_\mu^{i}\big([\bar c c] + [\bar q q]\big)$}
\label{sec:molecule1}

%
\begin{figure*}[hbt]
\begin{center}
\includegraphics[width=0.8\textwidth]{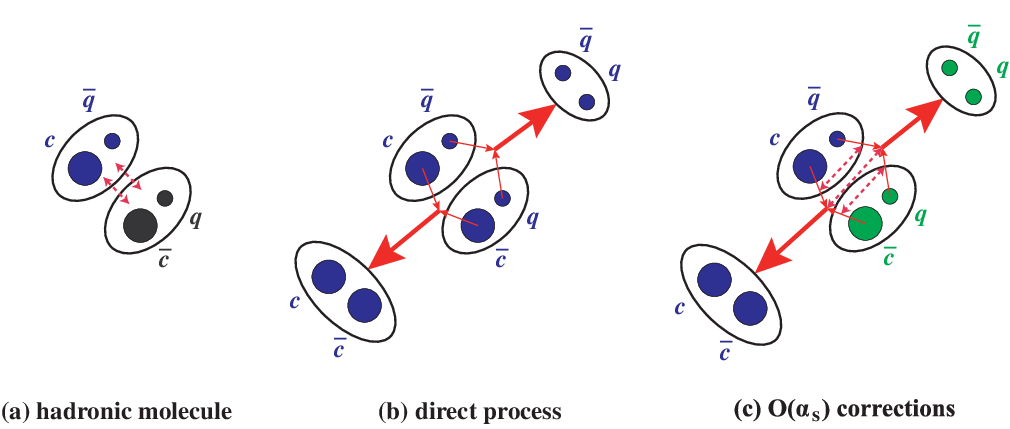}
\caption{The decay of a hadronic molecular state into one charmonium meson and one light meson, which can happen through either (b) a direct fall-apart process, or (c) a process with gluons exchanged.}
\label{fig:c}
\end{center}
\end{figure*}
%

As depicted in Fig.~\ref{fig:c}, when the $q$ and $\bar q$ quarks meet each other and the $c$ and $\bar c$ quarks meet each other at the same time, a hadronic molecular state decays into one charmonium meson and one light meson. This process can be described by the transformation~(\ref{eq:fierz4}):
\begin{eqnarray}
&& \xi^{\mathcal X}_\mu(x,y; I = 0)
\label{tran:molecule1}
\\ \nonumber &\Longrightarrow& + {1\over6}~\theta_\mu^1(x^\prime,y^\prime; I = 0) - {1\over6}~\theta_\mu^2(x^\prime,y^\prime; I = 0)
\\ \nonumber &&                - {{\rm i}\over6}~\theta_\mu^3(x^\prime,y^\prime; I = 0) + {{\rm i}\over6}~\theta_\mu^4(x^\prime,y^\prime; I = 0) + \cdots
\\ \nonumber &=& + {1\over6}~I^{S}(x^\prime)~P^{A}_\mu(y^\prime) - {1\over6}~I^{A}_\mu(x^\prime)~P^{S}(y^\prime)
\\ \nonumber &&  + {1\over12}~\epsilon_{\mu\nu\rho\sigma}~I^{V,\nu}(x^\prime)~P^{T,\rho\sigma}(y^\prime)
\\ \nonumber &&  - {1\over12}~\epsilon_{\mu\nu\rho\sigma}~I^{T,\rho\sigma}(x^\prime)~P^{V,\nu}(y^\prime)~+~\cdots \, .
\end{eqnarray}
Here we keep only the direct fall-apart process described by $\theta_\mu^{1,2,3,4}$, but neglect the $\mathcal{O}(\alpha_s)$ corrections described by $\theta_\mu^{5,6,7,8}$.

We repeat the same procedures as those done in Sec.~\ref{sec:diquark1}, and extract the following coupling constants from this transformation:
\begin{eqnarray}
\nonumber h_{\eta_c f_0} &\sim& - {\rm i} c_1~1.26 \times 10^{7}~{\rm MeV}^3 \, ,
\\ \nonumber h_{\chi_{c1} f_0} &\sim& - c_1~0.38 \times 10^{11}~{\rm MeV}^4 \, ,
\\ h^A_{\psi \omega} &=& - {\rm i} c_4~3.43 \times 10^{7}~{\rm MeV}^3 \, ,
\\ \nonumber h^B_{\psi \omega} &=& - {\rm i} c_4~1.16 \times 10^{7}~{\rm MeV}^3 \, ,
\\ \nonumber h_{\psi h_1} &=& + {\rm i} c_4~1.94 \times 10^{7}~{\rm MeV}^3 \, .
\end{eqnarray}
They are defined for the $| D \bar D^*; 1^{++} \rangle \rightarrow \eta_c f_0$, $\chi_{c1} f_0$, $J/\psi \omega$, and $J/\psi h_1$ decays. They all contain an overall factor $c_4$, which is related to the coupling of $\xi^{\mathcal X}_\mu(x,y)$ to the $X(3872)$ as well as the dynamical process $(x,y) \Longrightarrow (x^\prime, y^\prime)$ shown in Fig.~\ref{fig:c}.

Using these coupling constants, we further obtain
\begin{eqnarray}
{\mathcal{B}(| D \bar D^*; 1^{++} \rangle \rightarrow \eta_c f_0 \rightarrow \eta_c \pi \pi)
\over
\mathcal{B}(| D \bar D^*; 1^{++} \rangle \rightarrow J/\psi \omega \rightarrow J/\psi \pi \pi \pi)} &\sim& 0.091 \, ,
\label{eq:moleculekbr1}
\\ \nonumber {\mathcal{B}(| D \bar D^*; 1^{++} \rangle \rightarrow \chi_{c1} f_0 \rightarrow \chi_{c1} \pi \pi)
\over
\mathcal{B}(| D \bar D^*; 1^{++} \rangle \rightarrow J/\psi \omega \rightarrow J/\psi \pi \pi \pi)} &\sim& 0.086 \, ,
\\ \nonumber {\mathcal{B}(| D \bar D^*; 1^{++} \rangle \rightarrow J/\psi h_1 \rightarrow J/\psi \pi \pi \pi)
\over
\mathcal{B}(| D \bar D^*; 1^{++} \rangle \rightarrow J/\psi \omega \rightarrow J/\psi \pi \pi \pi)} &=& 1.4 \times 10^{-3} \, ,
\end{eqnarray}
which ratios are the same as Eqs.~(\ref{eq:diquarkbr1}), obtained in Sec.~\ref{sec:diquark1} for the compact tetraquark state $|0_{qc}1_{\bar q \bar c} ; 1^{++} \rangle$.

\subsection{$\xi^{\mathcal X}_\mu\big([\bar c q][\bar q c]\big) \longrightarrow \xi^i_\mu\big([\bar c q] + [\bar q c]\big)$}
\label{sec:molecule2}

Assuming the $X(3872)$ to be the $D \bar D^*$ hadronic molecular state of $J^{PC} = 1^{++}$, it can naturally decay into the $D \bar D^*$ final state, which fall-apart process can be described by itself:
\begin{eqnarray}
\xi^{\mathcal Z}_\mu(x,y; I=0) &\Longrightarrow& \xi_\mu^1(x^\prime,y^\prime; I=0)
\label{tran:molecule2}
\\ \nonumber &=& -{\rm i}~O^{V}_\mu(x^\prime) ~ O^{P}(y^\prime) + c.c. \, ,
\end{eqnarray}
and so
\begin{eqnarray}
\nonumber \langle X(p,\epsilon) | D^0(p_1) \bar D^{*0}(p_2,\epsilon_2) \rangle &\approx& -{{\rm i}c_5\over\sqrt2}~\lambda_D m_{D^*} f_{D^*}~\epsilon \cdot \epsilon_2
\\ &\equiv& h_{D \bar D^*}~\epsilon \cdot \epsilon_2 \, ,
\end{eqnarray}
where $c_5$ is an overall factor. Again, this decay might be kinematically forbidden, but the $| D \bar D^*; 1^{++} \rangle \rightarrow D^0 \bar D^{*0} + D^{*0} \bar D^0 \rightarrow D^0 \bar D^0 \pi^0$ process is surely kinematically allowed.

Numerically, we obtain
\begin{eqnarray}
h^S_{D \bar D^{*}} &=& - {\rm i}c_5~2.1 \times 10^{11}~{\rm MeV}^4 \, ,
\\ \nonumber h^D_{D \bar D^{*}} &=& 0 \, .
\end{eqnarray}
Comparing this decay with the $| D \bar D^*; 1^{++} \rangle \rightarrow J/\psi \omega \rightarrow J/\psi \pi \pi \pi$ decay studied in the previous subsection, we further obtain
\begin{eqnarray}
\nonumber && {\mathcal{B}(| D \bar D^*; 1^{++} \rangle \rightarrow D^0 \bar D^{*0} + D^{*0} \bar D^0 \rightarrow D^0 \bar D^0 \pi^0)
\over
\mathcal{B}(| D \bar D^*; 1^{++} \rangle \rightarrow J/\psi \omega \rightarrow J/\psi \pi \pi \pi)}
\\ &=& 4.5 \times {c_5^2 \over c_4^2} \, .
\end{eqnarray}
Therefore, $| D \bar D^*; 1^{++} \rangle$ mainly decays into two charmed mesons, because $c_5$ is probably larger than $c_4$.

\section{Isospin of the $X(3872)$}
\label{sec:isospin}

The isospin breaking effect of the $X(3872)$ is significant and important to understand its nature~\cite{Liu:2019zoy,Lebed:2016hpi,Esposito:2016noz,Guo:2017jvc,Ali:2017jda,Olsen:2017bmm,Karliner:2017qhf,Brambilla:2019esw}. As proposed in Ref.~\cite{Gamermann:2009fv}, this can be simply because the close proximity of the mass of the isoscalar $X(3872)$ to the neutral $D^0 \bar D^{*0}$ threshold. We argue that in this case the $X(3872)$ and the $D^0 \bar D^{*0}$ threshold together can be considered as a ``mixed'' state, not purely isoscalar any more, through which we can investigate the isovector decay channels of the $X(3872)$.

In this section we shall investigate the isospin breaking effect of the $X(3872)$ by freely choosing its quark content~\cite{Maiani:2004vq,Navarra:2006nd,Matheus:2006xi}, for example,
\begin{eqnarray}
\eta^1_\mu(\theta/\theta^\prime) &=& \cos\theta~\eta^1_\mu([uc][\bar u \bar c]) + \sin\theta~\eta^1_\mu([dc][\bar d \bar c])
\\ \nonumber &=& \cos\theta \times \left( u_a^{\rm T}\mathbb{C}\gamma_\mu c_b ~ \bar u_{a} \gamma_5\mathbb{C}\bar c_{b}^{\rm T} + \{ \gamma_\mu \leftrightarrow \gamma_5 \} \right)
\\ \nonumber && +~\sin\theta \times \{ u/\bar u \rightarrow d/\bar d \}
\\ \nonumber &\Rightarrow& \cos\theta^\prime~\eta^1_\mu(I=0) + \sin\theta^\prime~\eta^1_\mu(I=1) \, ,
\end{eqnarray}
where $\theta$ and $\theta^\prime$ are the two related mixing angles. We shall fine-tune them to be different from $\theta = 45^\circ/\theta^\prime = 0^\circ$, so that the $X(3872)$ is assumed not to be a purely isoscalar state. We shall study this effect separately for the compact tetraquark and hadronic molecule scenarios in the following subsections.

\subsection{Isospin breaking effect of $|0_{qc}1_{\bar q \bar c} ; 1^{++} \rangle$}

We repeat the same procedures as those done in Sec.~\ref{sec:diquark1}, and transform $\eta^{\mathcal X}_\mu(x,y;\theta_1^\prime)$ to $\theta_\mu^i(x,y;\theta_1^\prime)$ according to Eq.~(\ref{eq:fierz1}), from which we extract the following isovector decay channels:
\begin{enumerate}

\item The decay of $|0_{qc}1_{\bar q \bar c} ; 1^{++} \rangle$ into $\chi_{c0} \pi$ is contributed by $I^{S} \times J^{A}_\mu$:
\begin{eqnarray}
&& \langle X(p,\epsilon) | \chi_{c0}(p_1)~\pi(p_2) \rangle
\\ \nonumber &\approx& - {{\rm i}c_1\sin\theta_1^\prime\over3}~m_{\chi_{c0}} f_{\chi_{c0}} f_{\pi}~\epsilon \cdot p_2 \equiv g_{\chi_{c0} \pi}~\epsilon \cdot p_2 \, .
\end{eqnarray}
This process is kinematically allowed.

\item The decay of $|0_{qc}1_{\bar q \bar c} ; 1^{++} \rangle$ into $J/\psi \rho$ is contributed by both $I^{V,\nu} \times J^{T,\rho\sigma}$ and $I^{T,\rho\sigma} \times J^{V,\nu}$:
\begin{eqnarray}
&& \langle X(p,\epsilon) | J/\psi(p_1,\epsilon_1)~\rho(p_2,\epsilon_2) \rangle
\\ \nonumber &\approx& -{{\rm i}c_1\sin\theta_1^\prime\over3}~\epsilon_{\mu\nu\rho\sigma} \epsilon^\mu \epsilon_1^\nu \epsilon_2^\rho p_2^\sigma~m_{J/\psi}f_{J/\psi}f_{\rho}^T
\\ \nonumber &&        -{{\rm i}c_1\sin\theta_1^\prime\over3}~\epsilon_{\mu\nu\rho\sigma} \epsilon^\mu \epsilon_1^\nu \epsilon_2^\rho  p_1^\sigma~m_{\rho}f_{\rho}f_{J/\psi}^T
\\ \nonumber &\equiv& g^A_{\psi \rho}~\epsilon_{\mu\nu\rho\sigma} \epsilon^\mu \epsilon_1^\nu \epsilon_2^\rho p_2^\sigma + g^B_{\psi \rho}~\epsilon_{\mu\nu\rho\sigma} \epsilon^\mu \epsilon_1^\nu \epsilon_2^\rho p_1^\sigma \, .
\end{eqnarray}
If we use $m_{\rho^0} = 775.26$~MeV~\cite{pdg}, this decay would be kinematically forbidden, but the $|0_{qc}1_{\bar q \bar c} ; 1^{++} \rangle \rightarrow J/\psi \rho \rightarrow J/\psi \pi \pi$ process is surely kinematically allowed.

\end{enumerate}
Numerically, we obtain
\begin{eqnarray}
\nonumber g_{\chi_{c0} \pi} &=& -{\rm i}c_1\sin\theta_1^\prime~5.08 \times 10^{7}~{\rm MeV}^3 \, ,
\\ g^A_{\psi \rho} &=& - {\rm i}c_1\sin\theta_1^\prime~6.86 \times 10^{7}~{\rm MeV}^3 \, ,
\\ \nonumber g^B_{\psi \rho} &=& - {\rm i}c_1\sin\theta_1^\prime~2.29 \times 10^{7}~{\rm MeV}^3 \, .
\end{eqnarray}
Comparing these decays with the $|0_{qc}1_{\bar q \bar c} ; 1^{++} \rangle \rightarrow J/\psi \omega \rightarrow J/\psi \pi \pi \pi$ decay studied in Sec.~\ref{sec:diquark1} (a factor $\cos^2\theta_1^{\prime}$ needs to be multiplied there), we can use $\theta_1^\prime = \pm15^\circ$ to obtain
\begin{eqnarray}
\nonumber {\mathcal{B}(|0_{qc}1_{\bar q \bar c} ; 1^{++} \rangle \rightarrow \chi_{c0} \pi)
\over
\mathcal{B}(|0_{qc}1_{\bar q \bar c} ; 1^{++} \rangle \rightarrow J/\psi \rho \rightarrow J/\psi \pi \pi)} &=& 0.024 \, ,
\\ {\mathcal{B}(|0_{qc}1_{\bar q \bar c} ; 1^{++} \rangle \to J/\psi \omega \to J/\psi \pi \pi \pi)
\over
\mathcal{B}(|0_{qc}1_{\bar q \bar c} ; 1^{++} \rangle \rightarrow J/\psi \rho \rightarrow J/\psi \pi \pi)} &=& 1.6 \, ,
\label{eq:isospindiquark}
\end{eqnarray}
where the latter ratio has been fine-tuned to be the same as the recent BESIII experiment~\cite{Ablikim:2019zio}.

The isospin breaking effect can also affect the branching ratio of the $|0_{qc}1_{\bar q \bar c} ; 1^{++} \rangle \rightarrow D^0 \bar D^{*0} + D^{*0} \bar D^0 \rightarrow D^0 \bar D^0 \pi^0$ decay. Using $\theta_1^\prime = +15^\circ$, we obtain
\begin{eqnarray}
\nonumber && {\mathcal{B}(|0_{qc}1_{\bar q \bar c} ; 1^{++} \rangle \rightarrow D^0 \bar D^{*0} + D^{*0} \bar D^0 \rightarrow D^0 \bar D^0 \pi^0)
\over
\mathcal{B}(|0_{qc}1_{\bar q \bar c} ; 1^{++} \rangle \rightarrow J/\psi \omega \rightarrow J/\psi \pi \pi \pi)}
\\ &=& 0.52 \times {c_2^2 \over c_1^2} \, ,
\end{eqnarray}
while using $\theta_1^\prime = -15^\circ$, this ratio is calculated to be around $0.17\times{c_2^2 / c_1^2}$.

\subsection{Isospin breaking effect of $| D \bar D^*; 1^{++} \rangle$}

We repeat the same procedures as those done in Sec.~\ref{sec:molecule1}, and transform $\xi^{\mathcal X}_\mu(x,y;\theta_2^\prime)$ to $\theta_\mu^i(x,y;\theta_2^\prime)$ according to Eq.~(\ref{eq:fierz4}), from which we extract the following isovector coupling constants:
\begin{eqnarray}
\nonumber h_{\chi_{c0} \pi} &=& +{\rm i}c_4\sin\theta_2^\prime~2.54 \times 10^{7}~{\rm MeV}^3 \, ,
\\ h^A_{\psi \rho} &=& - {\rm i}c_4\sin\theta_2^\prime~3.43 \times 10^{7}~{\rm MeV}^3 \, ,
\\ \nonumber h^B_{\psi \rho} &=& - {\rm i}c_4\sin\theta_2^\prime~1.14 \times 10^{7}~{\rm MeV}^3 \, .
\end{eqnarray}
They are defined for the $| D \bar D^*; 1^{++} \rangle \rightarrow \chi_{c0} \pi$ and $J/\psi \rho$ decays.

We can use the same angle $\theta_2^\prime = \pm15^\circ$ to obtain
\begin{eqnarray}
\nonumber {\mathcal{B}(| D \bar D^*; 1^{++} \rangle \rightarrow \chi_{c0} \pi)
\over
\mathcal{B}(| D \bar D^*; 1^{++} \rangle \rightarrow J/\psi \rho \rightarrow J/\psi \pi \pi)} &=& 0.024 \, ,
\\ {\mathcal{B}(| D \bar D^*; 1^{++} \rangle \to J/\psi \omega \to J/\psi \pi \pi \pi)
\over
\mathcal{B}(| D \bar D^*; 1^{++} \rangle \rightarrow J/\psi \rho \rightarrow J/\psi \pi \pi)} &=& 1.6 \, ,
\label{eq:isospinmolecule}
\end{eqnarray}
which ratios are the same as Eqs.~(\ref{eq:isospindiquark}), obtained in the previous subsection for the compact tetraquark state $|0_{qc}1_{\bar q \bar c} ; 1^{++} \rangle$.

Again, the isospin breaking effect can affect the branching ratio of the $| D \bar D^*; 1^{++} \rangle \rightarrow D^0 \bar D^{*0} + D^{*0} \bar D^0 \rightarrow D^0 \bar D^0 \pi^0$ decay. Using $\theta_2^\prime = +15^\circ$, we obtain
\begin{eqnarray}
\nonumber && {\mathcal{B}(| D \bar D^*; 1^{++} \rangle \rightarrow D^0 \bar D^{*0} + D^{*0} \bar D^0 \rightarrow D^0 \bar D^0 \pi^0)
\over
\mathcal{B}(| D \bar D^*; 1^{++} \rangle \rightarrow J/\psi \omega \rightarrow J/\psi \pi \pi \pi)}
\\ &=& 7.4 \times {c_5^2 \over c_4^2} \, ,
\end{eqnarray}
while using $\theta_2^\prime = -15^\circ$, this ratio is calculated to be around $2.4\times{c_5^2 / c_4^2}$.

\section{Summary and discussions}
\label{sec:summary}

\begin{table*}[hbt]
\begin{center}
\renewcommand{\arraystretch}{1.5}
\caption{Relative branching ratios of the $X(3872)$ evaluated through the Fierz rearrangement. $\theta_{1,2}^\prime$ are the two angles related to the isospin breaking effect, which are fine-tuned to be $\theta^\prime_1 = \theta^\prime_2 = \pm15^{\rm o}$, so that ${\mathcal{B}(|0_{qc}1_{\bar q \bar c} ; 1^{++} \rangle \to J/\psi \omega \to J/\psi \pi \pi \pi) \over \mathcal{B}(|0_{qc}1_{\bar q \bar c} ; 1^{++} \rangle \rightarrow J/\psi \rho \rightarrow J/\psi \pi \pi)} = {\mathcal{B}(|D \bar D^*; 1^{++} \rangle \rightarrow J/\psi \omega \to J/\psi \pi \pi \pi) \over \mathcal{B}(|D \bar D^*; 1^{++} \rangle \rightarrow J/\psi \rho \rightarrow J/\psi \pi \pi)} = 1.6$~\cite{Ablikim:2019zio}.}
\begin{tabular}{ c | c | c | c | c | c | c}
\hline\hline
\multirow{2}{*}{Channels} & \multicolumn{3}{c|}{$|0_{qc}1_{\bar q \bar c}; 1^{++} \rangle$} & \multicolumn{3}{c}{$|D \bar D^*; 1^{++} \rangle$}
\\ \cline{2-7} & $I = 0/\theta_1^\prime = 0^{\rm o}$ &  $\theta_1^\prime = +15^{\rm o}$  &  $\theta_1^\prime = -15^{\rm o}$ & $I = 0/\theta_2^\prime = 0^{\rm o}$ &  $\theta_2^\prime = +15^{\rm o}$  &  $\theta_2^\prime = -15^{\rm o}$
\\ \hline\hline
${\mathcal{B}(X \rightarrow \eta_c f_0 \rightarrow \eta_c \pi \pi) \over \mathcal{B}(X \rightarrow J/\psi \omega \rightarrow J/\psi \pi \pi \pi)}$
& ~$\sim0.091$~ & ~$\sim0.091$~ & ~$\sim0.091$~ & ~$\sim0.091$~ & ~$\sim0.091$~ & ~$\sim0.091$~
\\ \hline
${\mathcal{B}(X \rightarrow \chi_{c1} f_0 \rightarrow \chi_{c1} \pi \pi) \over \mathcal{B}(X \rightarrow J/\psi \omega \rightarrow J/\psi \pi \pi \pi)}$
& ~$\sim0.086$~ & ~$\sim0.086$~ & ~$\sim0.086$~ & ~$\sim0.086$~ & ~$\sim0.086$~ & ~$\sim0.086$~
\\ \hline
${\mathcal{B}(X \rightarrow J/\psi h_1 \rightarrow J/\psi \pi \pi \pi) \over \mathcal{B}(X \rightarrow J/\psi \omega \rightarrow J/\psi \pi \pi \pi)}$
& ~$1.4 \times 10^{-3}$~ & ~$1.4 \times 10^{-3}$~ & ~$1.4 \times 10^{-3}$~ & ~$1.4 \times 10^{-3}$~ & ~$1.4 \times 10^{-3}$~ & ~$1.4 \times 10^{-3}$~
\\ \hline
${\mathcal{B}(X \rightarrow \chi_{c0} \pi) \over \mathcal{B}(X \rightarrow J/\psi \rho \rightarrow J/\psi \pi \pi)}$
& -- &  $0.024$  & $0.024$ & -- &  $0.024$  & $0.024$
\\ \hline
${\mathcal{B}(X \to J/\psi \omega \to J/\psi \pi \pi \pi) \over \mathcal{B}(X \rightarrow J/\psi \rho \rightarrow J/\psi \pi \pi)}$
& -- &  $1.6$~(input)  & $1.6$~(input) & -- &  $1.6$~(input)  & $1.6$~(input)
\\ \hline \hline
~${\mathcal{B}(X \rightarrow D^0 \bar D^{*0} + D^{*0} \bar D^0 \rightarrow D^0 \bar D^0 \pi^0) \over \mathcal{B}(X \rightarrow J/\psi \omega \rightarrow J/\psi \pi \pi \pi)}$~
& $0.32~t_1$ &  $0.52~t_1$ & $0.17~t_1$
& {$4.5~t_2$} & {$7.4~t_2$} & {$2.4~t_2$}
\\ \hline\hline
\end{tabular}
\label{tab:relative}
\end{center}
\end{table*}

In this paper we systematically construct the tetraquark currents of $J^{PC} = 1^{++}$ with the quark content $c \bar c q \bar q$ ($q=u/d$). We consider three configurations, $[cq][\bar c \bar q]$, $[\bar c q][\bar q c]$, and $[\bar c c][\bar q q]$, and we construct eight independent currents for each of them. Their relations are derived using the Fierz rearrangement of the Dirac and color indices, through which we study decay properties of the $X(3872)$:
\begin{itemize}

\item Based on the transformation of $[qc][\bar q \bar c] \to [\bar c c][\bar q q]$, we study decay properties of the $X(3872)$ as a compact tetraquark state into one charmonium meson and one light meson.

\item Based on the transformation of $[qc][\bar q \bar c] \to [\bar c q][\bar q c]$, we study decay properties of the $X(3872)$ as a compact tetraquark state into two charmed mesons.

\item Based on the transformation of the $[qc][\bar q \bar c]$ currents to the color-singlet-color-singlet $[\bar c c][\bar q q]$ and $[\bar c q][\bar q c]$ currents, we obtain the same relative branching ratios as those obtained using the above two transformations.

\item Based on the transformation of $[\bar c q][\bar q c] \to [\bar c c][\bar q q]$, we study decay properties of the $X(3872)$ as a hadronic molecular state into one charmonium meson and one light meson.

\item Based on the $[\bar c q][\bar q c]$ currents themselves, we study decay properties of the $X(3872)$ as a hadronic molecular state into two charmed mesons.

\end{itemize}

We first use isoscalar tetraquark currents to study decay properties of the $X(3872)$ as a purely isoscalar state, and then use isovector tetraquark currents to investigate its isospin breaking effect. The extracted relative branching ratios are summarized in Table~\ref{tab:relative}, where we have investigated the following interpretations of the $X(3872)$:
\begin{widetext}
\begin{itemize}

\item In the second, third, and fourth columns of Table~\ref{tab:relative}, $|0_{qc}1_{\bar q \bar c} ; 1^{++} \rangle$ denotes the compact tetraquark state of $J^{PC} = 1^{++}$, defined in Eq.~(\ref{def:diquark}). $|0_{qc}1_{\bar q \bar c} ; 1^{++} \rangle$ with $I=0$ ($\theta_1^\prime = 0^\circ$) is the purely isoscalar state, and those with $\theta_1^\prime = \pm15^\circ$ contain some isovector components. Using the mixing angle $\theta^\prime_1 = +15^{\rm o}$, we obtain
\begin{eqnarray}
\nonumber && {\mathcal{B}\left(|0_{qc}1_{\bar q \bar c}; 1^{++} \rangle \rightarrow
 J/\psi\omega (\rightarrow \pi \pi \pi)
: ~\,J/\psi\rho (\rightarrow \pi \pi)~\,
: ~\,\chi_{c0}\pi~
: \eta_c f_0 (\rightarrow \pi \pi)
: \chi_{c1} f_0 (\rightarrow \pi \pi)
: D^0 \bar D^{*0} (\rightarrow D^0 \bar D^{0} \pi^0)~
\right) \over \mathcal{B}\left(|0_{qc}1_{\bar q \bar c}; 1^{++} \rangle \rightarrow J/\psi\omega (\rightarrow \pi \pi \pi)\right)}
\\ &\sim&
~~~~~~~~~~~~~~~~~~~~~~~~~~~~~~~~~~~1~~~~~~~~ : ~~0.63~({\rm input})~~ : ~0.015~ : ~~0.091~(?)\,~ : ~~~0.086~(?)\,~~ : ~~~~~~~0.52~t_1~ \, ,
\end{eqnarray}
while using the mixing angle $\theta^\prime_1 = -15^{\rm o}$, we obtain
\begin{eqnarray}
\nonumber && {\mathcal{B}\left(|0_{qc}1_{\bar q \bar c}; 1^{++} \rangle \rightarrow
 J/\psi\omega (\rightarrow \pi \pi \pi)
: ~\,J/\psi\rho (\rightarrow \pi \pi)~\,
: ~\,\chi_{c0}\pi~
: \eta_c f_0 (\rightarrow \pi \pi)
: \chi_{c1} f_0 (\rightarrow \pi \pi)
: D^0 \bar D^{*0} (\rightarrow D^0 \bar D^{0} \pi^0)~
\right) \over \mathcal{B}\left(|0_{qc}1_{\bar q \bar c}; 1^{++} \rangle \rightarrow J/\psi\omega (\rightarrow \pi \pi \pi)\right)}
\\ &\sim&
~~~~~~~~~~~~~~~~~~~~~~~~~~~~~~~~~~~1~~~~~~~~ : ~~0.63~({\rm input})~~ : ~0.015~ : ~~0.091~(?)\,~ : ~~~0.086~(?)\,~~ : ~~~~~~~0.17~t_1~ \, .
\end{eqnarray}

\item In the fifth, sixth, and seventh columns of Table~\ref{tab:relative}, $| D \bar D^*; 1^{++} \rangle$ denotes the hadronic molecular state of $J^{PC} = 1^{++}$, defined in Eq.~(\ref{def:molecule}). $| D \bar D^*; 1^{++} \rangle$ with $I=0$ ($\theta_2^\prime = 0^\circ$) is the purely isoscalar state, and those with $\theta_2^\prime = \pm15^\circ$ contain some isovector components. Using the mixing angle $\theta^\prime_2 = +15^{\rm o}$, we obtain
\begin{eqnarray}
\nonumber && {\mathcal{B}\left(|D \bar D^*; 1^{++} \rangle \rightarrow
 J/\psi\omega (\rightarrow \pi \pi \pi)
: ~\,J/\psi\rho (\rightarrow \pi \pi)~\,
: ~\chi_{c0}\pi~
: \eta_c f_0 (\rightarrow \pi \pi)
: \chi_{c1} f_0 (\rightarrow \pi \pi)
: D^0 \bar D^{*0} (\rightarrow D^0 \bar D^{0} \pi^0)~
\right) \over \mathcal{B}\left(|D \bar D^*; 1^{++} \rangle \rightarrow J/\psi\omega (\rightarrow \pi \pi \pi)\right)}
\\ &\sim&
~~~~~~~~~~~~~~~~~~~~~~~~~~~~~~~~~\,1~~~~~~~~ : ~~0.63~({\rm input})~~ : ~0.015~ : ~~0.091~(?)\,~ : ~~~0.086~(?)\,~~ : ~~~~~~~7.4~t_2~ \, ,
\end{eqnarray}
while using the mixing angle $\theta^\prime_2 = -15^{\rm o}$, we obtain
\begin{eqnarray}
\nonumber && {\mathcal{B}\left(|D \bar D^*; 1^{++} \rangle \rightarrow
 J/\psi\omega (\rightarrow \pi \pi \pi)
: ~\,J/\psi\rho (\rightarrow \pi \pi)~\,
: ~\chi_{c0}\pi~
: \eta_c f_0 (\rightarrow \pi \pi)
: \chi_{c1} f_0 (\rightarrow \pi \pi)
: D^0 \bar D^{*0} (\rightarrow D^0 \bar D^{0} \pi^0)~
\right) \over \mathcal{B}\left(|D \bar D^*; 1^{++} \rangle \rightarrow J/\psi\omega (\rightarrow \pi \pi \pi)\right)}
\\ &\sim&
~~~~~~~~~~~~~~~~~~~~~~~~~~~~~~~~~\,1~~~~~~~~ : ~~0.63~({\rm input})~~ : ~0.015~ : ~~0.091~(?)\,~ : ~~~0.086~(?)\,~~ : ~~~~~~~2.4~t_2~ \, .
\end{eqnarray}

\end{itemize}
\end{widetext}
In the above expressions, we define the ratio $t_1 \equiv {c_2^2 / c_1^2}$ to be the parameter measuring which process happens more easily, the process depicted in Fig.~\ref{fig:a}(b) or the process depicted in Fig.~\ref{fig:b}(b). Because the exchange of one light quark with another light quark seems to be easier than the exchange of one light quark with another heavy quark~\cite{Landau,Maiani:2017kyi}, it can be the case that $t_1 \geq 1$. As discussed in Sec.~\ref{sec:molecule2}, $c_5$ is probably larger than $c_4$, so that the other ratio $t_2 \equiv {c_5^2 / c_4^2} \geq 1$.

The above relative branching ratios extracted in the present study turn out to be very much different. This might be one of the reasons why many multiquark states were observed only in a few decay channels~\cite{Chen:2019wjd}. We note that in this paper we only consider the leading-order fall-apart decays described by color-singlet-color-singlet meson-meson currents, but neglect the $\mathcal{O}(\alpha_s)$ corrections described by color-octet-color-octet meson-meson currents, so there can be other possible decay channels, such as $X(3872) \rightarrow \chi_{c1} \pi$~\cite{Ablikim:2019soz}. Besides, there is still one parameter not considered in above analyses, that is the phase angle between $S$- and $D$-wave coupling constants. We shall investigate its relevant uncertainty in \ref{app:phase}.

Based on Table~\ref{tab:relative}, we conclude this paper. Generally speaking, compared to the $Z_c(3900)$ studied in Ref.~\cite{Chen:2019wjd}, the results of this paper suggest that decay channels of the $X(3872)$ are quite limited:
\begin{itemize}

\item The $X(3872)$ can couple to the $\chi_{c0} \eta$, $\chi_{c0} f_1(1285)$, and $h_c \omega$ channels, but all of them are kinematically forbidden.

\item The $X(3872)$ can couple to the isovector channels $J/\psi \rho$ and $\chi_{c0} \pi$, but both of them are due to the isospin breaking effect.

\item The $X(3872)$ can couple to the $D^0 \bar D^{*0}$ and $J/\psi \rho$ channels, but its mass is very close to the relevant thresholds. Hence, in the present study we calculate widths of the three-body decays $X \rightarrow D^0 \bar D^{*0} + D^{*0} \bar D^0 \rightarrow D^0 \bar D^0 \pi^0$ and $X \rightarrow J/\psi \rho \rightarrow J/\psi \pi \pi$.

\item The $X(3872)$ can couple to the $J/\psi \omega$ and $J/\psi h_1(1170)$ channels, but both of them are kinematically forbidden. Hence, in the present study we calculate widths of the four-body decays $X \rightarrow J/\psi \omega \rightarrow J/\psi \pi \pi \pi$ and $X \rightarrow J/\psi h_1 \rightarrow J/\psi \pi \pi \pi$.

\item The decay processes $X \rightarrow \eta_c f_0 \rightarrow \eta_c \pi \pi$ and $X \rightarrow \chi_{c1} f_0 \rightarrow \chi_{c1} \pi \pi$ might be possible. In this paper we simply use the $f_0(500)$ to estimate widths of these two processes, but note that the obtained results do significantly depend on the nature of light scalar mesons, which are still quite ambiguous~\cite{Pelaez:2015qba}.

\end{itemize}

To end this paper, we give several comments and proposals:
\begin{itemize}

\item The hadronic molecular state $| D \bar D^*; 1^{++} \rangle$ mainly decays into two charmed mesons, because $c_5$ is probably larger than $c_4$. The compact tetraquark state $|0_{qc}1_{\bar q \bar c} ; 1^{++} \rangle$ may also mainly decay into two charmed mesons after taking into account the barrier of the diquark-antidiquark potential (see detailed discussions in Ref.~\cite{Maiani:2017kyi} proposing $c_2 \gg c_1$).

\item The isospin breaking effect of the $X(3872)$ is significant and important to understand its nature~\cite{Liu:2019zoy,Lebed:2016hpi,Esposito:2016noz,Guo:2017jvc,Ali:2017jda,Olsen:2017bmm,Karliner:2017qhf,Brambilla:2019esw}. The isovector decay channel $X(3872) \rightarrow J/\psi \rho \rightarrow J/\psi \pi \pi$ has been well observed in experiments, and recently measured by the BESIII experiment~\cite{Ablikim:2019zio} to be:
    \begin{equation}
    {\mathcal{B}(X \to J/\psi \omega \to J/\psi \pi \pi \pi) \over \mathcal{B}(X \rightarrow J/\psi \rho \rightarrow J/\psi \pi \pi)} = 1.6^{+0.4}_{-0.3} \pm 0.2 \, .
    \end{equation}
    In the present study we can well reproduce this value under both the compact tetraquark and hadronic molecule interpretations.

    Besides this, our result suggests that there can be another isovector decay channel $X(3872) \rightarrow \chi_{c0} \pi$. Under both the compact tetraquark and hadronic molecule interpretations, we obtain
    \begin{equation}
    {\mathcal{B}(X \rightarrow \chi_{c0} \pi) \over \mathcal{B}(X \rightarrow J/\psi \rho \rightarrow J/\psi \pi \pi)} = 0.024 \, .
    \end{equation}
    We refer to Refs.~\cite{Dubynskiy:2007tj,Fleming:2008yn,Fleming:2011xa,Dong:2009yp,Guo:2010ak,Harada:2010bs,Mehen:2015efa,Zhou:2019swr} for more theoretical studies, and propose to study the $X(3872) \rightarrow \chi_{c0} \pi$ decay in the BESIII, Belle-II, and LHCb experiments to better understand the isospin breaking effect of the $X(3872)$.

\item Our result suggests that the decay processes $X(3872) \rightarrow \eta_c f_0 \rightarrow \eta_c \pi \pi$ and $X(3872) \rightarrow \chi_{c1} f_0 \rightarrow \chi_{c1} \pi \pi$ might be possible. We note that light scalar mesons have a complicated nature, so our results on these processes are just roughly estimations.

    We notice that the BaBar experiment~\cite{Lees:2012me} did not observe the $\gamma\gamma \to X(3872) \to \eta_c \pi \pi$ process, but that experiment was performed after assuming $X(3872)$ to be a spin-2 state. Moreover, there seems to be a dip structure just at the mass of the $X(3872)$ in the $\eta_c \pi \pi$ invariant mass spectrum, as shown in Fig.~6(f) of Ref.~\cite{Lees:2012me}. We also notice that the Belle experiment~\cite{Bhardwaj:2015rju} did not observe the $X(3872) \to \chi_{c1} \pi \pi$ decay. They extracted the following upper limit
    \begin{equation}
    \nonumber {\mathcal{B}(B^+ \to K^+ X) \mathcal{B}(X \to \chi_{c1} \pi \pi)} < 1.5 \times 10^{-6} \, ,
    \end{equation}
    at 90\% C.L. Together with another Belle experiment~\cite{Kato:2017gfv} measuring
    \begin{equation}
    \nonumber \mathcal{B}(B^+ \to K^+ X) = (1.2 \pm 1.1 \pm 0.1) \times 10^{-4} \, ,
    \end{equation}
    one may roughly estimate
    \begin{equation}
    \mathcal{B}(X \to \chi_{c1} \pi \pi) < 1.3 \times 10^{-2} \, ,
    \end{equation}
    which value seems not small enough to rule out the $X(3872) \to \chi_{c1} \pi \pi$ decay channel.

    Again, we refer to Refs.~\cite{Dubynskiy:2007tj,Fleming:2008yn,Fleming:2011xa,Dong:2009yp,Guo:2010ak,Harada:2010bs,Mehen:2015efa,Zhou:2019swr,Olsen:2004fp} for more discussions, and propose to reanalysis the $X(3872) \rightarrow \eta_c f_0 \rightarrow \eta_c \pi \pi$ and $X(3872) \to \chi_{c1} f_0 \to \chi_{c1} \pi \pi$ processes in the BESIII, Belle-II, and LHCb experiments to search for more decay channels of the $X(3872)$.

\end{itemize}

%
\section*{Acknowledgments}
%

This project is supported by
the National Natural Science Foundation of China under Grant No.~11722540 and No.~12075019,
the Jiangsu Provincial Double-Innovation Program under Grant No.~JSSCRC2021488,
and
the Fundamental Research Funds for the Central Universities.

\appendix

\section{Formulae of decay amplitudes and decay widths}
\label{sec:width}

In this appendix we give formulae of decay amplitudes and decay widths used in the present study. Especially, the mass of the $X(3872)$ is taken from PDG~\cite{pdg} to be
\begin{equation}
m_X = 3871.69 ~{\rm MeV} \, .
\end{equation}

\subsection{Two-body decay $X \to \chi_{c0} \pi^0$}

The decay amplitude of the two-body decay $X(3872) \to \chi_{c0} \pi^0$ is
\begin{equation}
\mathcal{M} \left( X(\epsilon,p) \to \chi_{c0}(p_1) \pi^0(p_2) \right) = g_{\chi_{c0} \pi} ~ \epsilon \cdot p_2 \, .
\end{equation}
This amplitude can be used to evaluate its decay width:
\begin{equation}
\Gamma \left( X \to \chi_{c0} \pi^0 \right) = { \left| \vec p_2 \right| \over 8 \pi m_X^2} \left|g^2_{\chi_{c0} \pi}\right| {p_2^\mu p_2^\nu\over3} \left( g_{\mu\nu} - {p_{\mu} p_{\nu} \over m_X^2} \right) \, ,
\end{equation}
where we have used the following formula for the vector meson
\begin{equation}
\sum \epsilon_{\mu} \epsilon^*_{\nu} = g_{\mu\nu} - {p_{\mu} p_{\nu} \over m_X^2} \, .
\end{equation}

\subsection{Three-body decay $X \to J/\psi \rho^0 \rightarrow J/\psi \pi^+ \pi^-$}

First we need to investigate the two-body decay $\rho^0 \rightarrow \pi^+ \pi^-$, whose amplitude is
\begin{equation}
\mathcal{M} \left( \rho^0(\epsilon,p) \rightarrow \pi^+(p_1) \pi^-(p_2) \right) = g_{\rho\pi\pi} ~ \epsilon \cdot \left( p_1 - p_2 \right) \, ,
\end{equation}
so that
\begin{eqnarray}
\nonumber \Gamma \left( \rho^0 \rightarrow \pi^+ \pi^- \right) &=& {1\over3}{ \left| \vec p_2 \right| \over 8 \pi m_\rho^2} \left|g^2_{\rho\pi\pi}\right| \left( g_{\mu\nu} - {p_{\mu} p_{\nu} \over m_\rho^2} \right)
\\ && \times~{\left(p_1^\mu - p_2^\mu\right) \left(p_1^\nu - p_2^\nu\right)}  \, .
\end{eqnarray}
We can use the experimental parameters $\Gamma_{\rho^0} = 147.8$~MeV and $\mathcal{B}({\rho^0 \to \pi^+ \pi^-}) \approx 100\%$~\cite{pdg} to extract
\begin{equation}
g_{\rho\pi\pi} = 5.94 \, .
\end{equation}

The decay amplitude of the three-body decay $X(3872) \to J/\psi \rho^0 \rightarrow J/\psi \pi^+ \pi^-$ is
\begin{eqnarray}
&& \mathcal{M} \Big( X(\epsilon,p) \to J/\psi(\epsilon_1,p_1) \rho^0(\epsilon^\prime,q)
\\ \nonumber && ~~~~~~~~~~~~~~~ \rightarrow J/\psi(\epsilon_1,p_1) \pi^+(p_2) \pi^-(p_3) \Big)
\\ \nonumber &=& g_{\rho\pi\pi} ~ \left( g^A_{\psi \rho}~\epsilon_{\mu\nu\rho\sigma} \epsilon^\mu \epsilon_1^\nu q^\sigma + g^B_{\psi \rho}~\epsilon_{\mu\nu\rho\sigma} \epsilon^\mu \epsilon_1^\nu p_1^\sigma \right)
\\ \nonumber && \times~{ p_{2,\alpha} - p_{3,\alpha} \over q^2 - m_\rho^2 + {\rm i} m_\rho \Gamma_\rho } ~ \left( g^{\rho\alpha} - {q^{\rho} q^{\alpha} \over m_\rho^2} \right) \, .
\end{eqnarray}
This amplitude can be used to evaluate its decay width:
\begin{eqnarray}
&& \Gamma \left( X \to J/\psi \rho^0 \rightarrow J/\psi \pi^+ \pi^- \right)
\\ \nonumber &=& {1\over3}{1\over(2\pi)^3}{ g_{\rho\pi\pi}^2 \over 32 m_X^3} \int {\rm d}m_{12}^2 {\rm d}m_{23}^2 ~ \left|{1 \over q^2 - m_\rho^2 + {\rm i} m_\rho \Gamma_\rho }\right|^2
\\ \nonumber && \times \left( g^A_{\psi \rho}~\epsilon_{\mu\nu\rho\sigma} q^\sigma + g^B_{\psi \rho}~\epsilon_{\mu\nu\rho\sigma} p_1^\sigma \right)
\\ \nonumber && \times \left( g^{A*}_{\psi \rho}~\epsilon_{\mu^\prime\nu^\prime\rho^\prime\sigma^\prime} q^{\sigma^\prime} + g^{B*}_{\psi \rho}~\epsilon_{\mu^\prime\nu^\prime\rho^\prime\sigma^\prime} p_1^{\sigma^\prime} \right)
\\ \nonumber && \times \left( g^{\mu\mu^\prime} - {p^{\mu} p^{\mu^\prime} \over m_X^2} \right) ~ \left( g^{\nu\nu^\prime} - {p_{1}^\nu p_{1}^{\nu^\prime} \over m_{J/\psi}^2} \right)
\\ \nonumber && \times \left( g^{\rho\alpha} - {q^{\rho} q^{\alpha} \over m_\rho^2} \right) ~ \left( g^{\rho^\prime\alpha^\prime} - {q^{\rho^\prime} q^{\alpha^\prime} \over m_\rho^2} \right)
\\ \nonumber && \times \left( p_{2,\alpha} - p_{3,\alpha} \right) ~ \left( p_{2,\alpha^\prime} - p_{3,\alpha^\prime} \right) \, .
\end{eqnarray}

\subsection{Three-body decay $X \to \eta_c f_0 \rightarrow \eta_c \pi \pi$}

First we need to investigate the two-body decay $f_0(500) \rightarrow \pi \pi$, whose amplitude is
\begin{equation}
\mathcal{M} \left( f_0(p) \rightarrow \pi(p_1) \pi(p_2) \right) = g_{f_0\pi\pi} \, .
\end{equation}
In this case we do not differentiate $\pi^{\pm,0}$. The above amplitude can be used to evaluate its decay width:
\begin{eqnarray}
\nonumber \Gamma \left( f_0 \rightarrow \pi \pi \right) &=& { \left| \vec p_2 \right| \over 8 \pi m_{f_0}^2}~\left|g^2_{f_0\pi\pi}\right|  \, .
\end{eqnarray}
We can use the experimental parameters $m_{f_0} = 512$~MeV and $\Gamma_{f_0} = 376$~MeV~\cite{Ablikim:2016frj} to extract
\begin{equation}
g_{f_0\pi\pi} = 3380~{\rm MeV} \, .
\end{equation}

The decay amplitude of the three-body decay $X(3872) \to \eta_c f_0 \rightarrow \eta_c \pi \pi$ is
\begin{eqnarray}
\nonumber && \mathcal{M} \Big( X(\epsilon,p) \to \eta_c(p_1) f_0(q) \rightarrow \eta_c(p_1) \pi(p_2) \pi(p_3) \Big)
\\ &=& g_{f_0\pi\pi} ~ g_{\eta_c f_0}~{ \epsilon \cdot p_1 \over q^2 - m_{f_0}^2 + {\rm i} m_{f_0} \Gamma_{f_0} } \, .
\end{eqnarray}
This amplitude can be used to evaluate its decay width:
\begin{eqnarray}
&& \Gamma \left( X \to \eta_c f_0 \rightarrow \eta_c \pi \pi \right)
\\ \nonumber &=& {1\over3}{1\over(2\pi)^3}{ g_{f_0\pi\pi}^2 \over 32 m_X^3} \int {\rm d}m_{12}^2 {\rm d}m_{23}^2 ~ \left|{1 \over q^2 - m_{f_0}^2 + {\rm i} m_{f_0} \Gamma_{f_0} }\right|^2
\\ \nonumber && \times ~ g_{\eta_c f_0} g^*_{\eta_c f_0}~ \left( g^{\mu\nu} - {p^{\mu} p^{\nu} \over m_X^2} \right) ~ p_{1,\mu}p_{1,\nu} \, .
\end{eqnarray}

The $X(3872) \to \chi_{c1} f_0 \rightarrow \chi_{c1} \pi \pi$ decay can be similarly studied.

\subsection{Three-body decay $X \rightarrow D^0 \bar D^{*0} \rightarrow D^0 \bar D^0 \pi^0$}

First we need to investigate the two-body decay $D^{*0} \rightarrow D^0 \pi^0$, whose amplitude is
\begin{equation}
\mathcal{M} \left( D^{*0}(\epsilon,p) \rightarrow D^0(p_1) \pi^0(p_2) \right) = g_{D^{*} D \pi} ~ \epsilon \cdot p_2 \, ,
\end{equation}
so that
\begin{equation}
\Gamma \left( D^{*0} \rightarrow D^0 \pi^0 \right) = { \left| \vec p_2 \right|  \left|g^2_{D^{*} D \pi}\right| \over 8 \pi m_{D^{*}}^2} { p_2^\mu p_2^\nu\over3}
 \left( g_{\mu\nu} - {p_{\mu} p_{\nu} \over m_{D^{*}}^2} \right) \, .
\end{equation}
We can use the parameters $\Gamma_{D^{*0}} = 83.3$~keV~\cite{Rosner:2013sha} and $\mathcal{B}({D^{*0} \rightarrow D^0 \pi^0}) = 64.7\%$~\cite{pdg} to extract
\begin{equation}
g_{D^{*} D \pi} = 14.6 \, .
\end{equation}

The decay amplitude of the three-body decay $X(3872) \rightarrow D^0 \bar D^{*0} \rightarrow D^0 \bar D^0 \pi^0$ is
\begin{eqnarray}
&& \mathcal{M} \Big( X(\epsilon,p) \to D^0(p_1) \bar D^{*0}(\epsilon^\prime,q)
\\ \nonumber && ~~~~~~~~~~~~~~~ \rightarrow D^0(p_1) \bar D^0(p_2) \pi^0(p_3) \Big)
\\ \nonumber &=& g_{D^{*} D \pi}~\left( g^S_{D \bar D^{*}}~\epsilon_\mu + g^D_{D \bar D^{*}}~(\epsilon \cdot q ~ p_{1,\mu} - p_1 \cdot q~\epsilon_\mu) \right)
\\ \nonumber && \times~{ p_{3,\nu} \over q^2 - m_{D^{*}}^2 + {\rm i} m_{D^{*}} \Gamma_{D^{*}} } ~ \left( g^{\mu\nu} - {q^{\mu} q^{\nu} \over m_{D^{*}}^2} \right) \, .
\end{eqnarray}
This amplitude can be used to evaluate its decay width:
\begin{eqnarray}
&& \Gamma \left( X \rightarrow D^0 \bar D^{*0} \rightarrow D^0 \bar D^0 \pi^0 \right)
\\ \nonumber &=& {1\over3}{1\over(2\pi)^3}{ g_{D^{*} D \pi}^2 \over 32 m_X^3} \int {\rm d}m_{12}^2 {\rm d}m_{23}^2
\\ \nonumber && \times \left|{1 \over q^2 - m_{D^{*}}^2 + {\rm i} m_{D^{*}} \Gamma_{D^{*}} }\right|^2~\left( g^{\mu\mu^\prime} - {p^{\mu} p^{\mu^\prime} \over m_X^2} \right)
\\ \nonumber && \times \left( g^S_{D \bar D^{*}}~g_{\mu\nu} + g^D_{D \bar D^{*}}~(p_{1,\nu}q_\mu - p_1 \cdot q~g_{\mu\nu}) \right)
\\ \nonumber && \times \left( g^{S*}_{D \bar D^{*}}~g_{\mu^\prime\nu^\prime} + g^{D*}_{D \bar D^{*}}~(p_{1,\nu^\prime}q_{\mu^\prime} - p_1 \cdot q~g_{\mu^\prime\nu^\prime}) \right)
\\ \nonumber && \times \left( g^{\nu\rho} - {q^{\nu} q^{\rho} \over m_{D^{*}}^2} \right) ~ \left( g^{\nu^\prime\rho^\prime} - {q^{\nu^\prime} q^{\rho^\prime} \over m_{D^{*}}^2} \right) ~ p_{3,\rho} p_{3,\rho^\prime} \, .
\end{eqnarray}
The width of the $X(3872) \rightarrow D^{*0} \bar D^0 \rightarrow D^0 \bar D^0 \pi^0$ decay is the same as the $X(3872) \rightarrow D^0 \bar D^{*0} \rightarrow D^0 \bar D^0 \pi^0$ decay, and we assume them to be non-coherent in the present study.

\subsection{Four-body decay $X \to J/\psi \omega \rightarrow J/\psi \pi^+ \pi^- \pi^0$}

First we need to investigate the three-body decay $\omega \rightarrow \pi^+ \pi^- \pi^0$, whose amplitude is
\begin{eqnarray}
&& \mathcal{M} \left( \omega(\epsilon,p) \rightarrow \pi^+(p_1) \pi^-(p_2) \pi^0(p_3) \right)
\\ \nonumber && ~~~~~~~~~~~~~~~~~~~~~~~~~~~~~~~~~~~ = g_{\omega3\pi} ~ \epsilon_{\mu\nu\rho\sigma} \epsilon^\mu p_1^\nu p_2^\rho p_3^\sigma \, ,
\end{eqnarray}
so that
\begin{eqnarray}
&& \Gamma \left( \omega \rightarrow \pi^+ \pi^- \pi^0 \right)
\\ \nonumber &=& {1\over3}{1\over(2\pi)^3}{ g_{\omega3\pi}^2 \over 32 m_\omega^3} \int {\rm d}m_{12}^2 {\rm d}m_{23}^2~\left( g^{\mu\mu^\prime} - {p^{\mu} p^{\mu^\prime} \over m_\omega^2} \right)
\\ \nonumber && \times~\epsilon_{\mu\nu\rho\sigma} p_1^\nu p_2^\rho p_3^\sigma~\epsilon_{\mu^\prime\nu^\prime\rho^\prime\sigma^\prime} p_1^{\nu^\prime} p_2^{\rho^\prime} p_3^{\sigma^\prime} \, .
\end{eqnarray}
We can use the experimental parameters $\Gamma_{\omega} = 8.49$~MeV and $\mathcal{B}({\omega \to \pi^+ \pi^- \pi^0}) = 89.3\%$~\cite{pdg} to extract
\begin{equation}
g_{\omega3\pi} = 1.4 \times 10^{-6}~{\rm MeV}^{-3} \, .
\end{equation}

The decay amplitude of the four-body decay $X(3872) \to J/\psi \omega \rightarrow J/\psi \pi^+ \pi^- \pi^0$ is
\begin{eqnarray}
&& \mathcal{M} \Big( X(\epsilon,p) \to J/\psi(\epsilon_1,p_1) \omega(\epsilon^\prime,q)
\\ \nonumber && ~~~~~~~~~~~~~~~ \rightarrow J/\psi(\epsilon_1,p_1) \pi^+(p_2) \pi^-(p_3) \pi^0(p_4) \Big)
\\ \nonumber &=& g_{\omega3\pi} ~ \left( g^A_{\psi \omega}~\epsilon_{\mu\nu\rho\sigma} \epsilon^\mu \epsilon_1^\nu q^\sigma + g^B_{\psi \omega}~\epsilon_{\mu\nu\rho\sigma} \epsilon^\mu \epsilon_1^\nu p_1^\sigma \right)
\\ \nonumber && \times~{ 1 \over q^2 - m_\omega^2 + {\rm i} m_\omega \Gamma_\omega }~\left( g^{\rho\alpha} - {q^{\rho} q^{\alpha} \over m_\omega^2} \right)~\epsilon_{\alpha\beta\gamma\zeta} p_2^\beta p_3^\gamma p_4^\zeta\, .
\end{eqnarray}
This amplitude can be used to evaluate its decay width:
\begin{eqnarray}
&& \Gamma \left( X \to J/\psi \omega \rightarrow J/\psi \pi^+ \pi^- \pi^0 \right)
\\ \nonumber &=& {g_{\omega3\pi}^2\over3}{ (2\pi)^4 \over 2 m_X} \int {{\rm d}^3p_1\over(2\pi)^3 2E_1} {{\rm d}^3p_2\over(2\pi)^3 2E_2} {4\pi p_{3x}^2\over(2\pi)^6 2E_3 2E_4}
\\ \nonumber && \times \left|{1 \over {p_{3x} \over E_3} + {p_{1x} + p_{2x} + p_{3x} \over E_4} }\right| ~ \left|{1 \over q^2 - m_\omega^2 + {\rm i} m_\omega \Gamma_\omega }\right|^2
\\ \nonumber && \times \left( g^A_{\psi \omega}~\epsilon_{\mu\nu\rho\sigma} q^\sigma + g^B_{\psi \omega}~\epsilon_{\mu\nu\rho\sigma} p_1^\sigma \right)
\\ \nonumber && \times \left( g^{A*}_{\psi \omega}~\epsilon_{\mu^\prime\nu^\prime\rho^\prime\sigma^\prime} q^{\sigma^\prime} + g^{B*}_{\psi \omega}~\epsilon_{\mu^\prime\nu^\prime\rho^\prime\sigma^\prime} p_1^{\sigma^\prime} \right)
\\ \nonumber && \times \left( g^{\mu\mu^\prime} - {p^{\mu} p^{\mu^\prime} \over m_X^2} \right) ~ \left( g^{\nu\nu^\prime} - {p_{1}^\nu p_{1}^{\nu^\prime} \over m_{J/\psi}^2} \right)
\\ \nonumber && \times \left( g^{\rho\alpha} - {q^{\rho} q^{\alpha} \over m_\omega^2} \right) ~ \left( g^{\rho^\prime\alpha^\prime} - {q^{\rho^\prime} q^{\alpha^\prime} \over m_\omega^2} \right)
\\ \nonumber && \times~\epsilon_{\alpha\beta\gamma\zeta} p_2^\beta p_3^\gamma p_4^\zeta ~ \epsilon_{\alpha^\prime\beta^\prime\gamma^\prime\zeta^\prime} p_2^{\beta^\prime} p_3^{\gamma^\prime} p_4^{\zeta^\prime} \, .
\end{eqnarray}
The phase space integration is done in the reference frame where $p_3 = \left( E_3, p_{3x}, 0, 0 \right)$, and $p_{3x}$ satisfies $p_{3x}>0$ as well as
\begin{equation}
E_1 + E_2 + E_3 + E_4 = m_X \, .
\end{equation}

\subsection{Four-body decay $X \to J/\psi h_1 \rightarrow J/\psi \pi^+ \pi^- \pi^0$}

First we need to investigate the three-body decay $h_1(1170) \rightarrow \rho \pi \rightarrow \pi^+ \pi^- \pi^0$, whose amplitude is simply assumed to be
\begin{equation}
\mathcal{M} \left( h_1(\epsilon,p) \rightarrow \pi^+(p_1) \pi^-(p_2) \pi^0(p_3) \right) = g_{h_13\pi} ~ \epsilon \cdot p_3 \, ,
\end{equation}
so that
\begin{eqnarray}
\nonumber \Gamma \left( h_1 \rightarrow \pi^+ \pi^- \pi^0 \right) &=& {1\over3}{1\over(2\pi)^3}{ g_{h_13\pi}^2 \over 32 m_{h_1}^3} \int {\rm d}m_{12}^2 {\rm d}m_{23}^2
\\ \nonumber && \times~p_{3,\mu}p_{3,\mu^\prime} ~ \left( g^{\mu\mu^\prime} - {p^{\mu} p^{\mu^\prime} \over m_{h_1}^2} \right) \, .
\end{eqnarray}
We can use the experimental parameters $\Gamma_{h_1} = 360$~MeV~\cite{pdg} to estimate
\begin{equation}
g_{h_13\pi} \approx 0.39~{\rm MeV}^{-1} \, .
\end{equation}

The decay amplitude of the four-body decay $X(3872) \to J/\psi h_1 \rightarrow J/\psi \pi^+ \pi^- \pi^0$ is
\begin{eqnarray}
&& \mathcal{M} \Big( X(\epsilon,p) \to J/\psi(\epsilon_1,p_1) h_1(\epsilon^\prime,q)
\\ \nonumber && ~~~~~~~~~~~~~~~ \rightarrow J/\psi(\epsilon_1,p_1) \pi^+(p_2) \pi^-(p_3) \pi^0(p_4) \Big)
\\ \nonumber &=& g_{h_13\pi} ~ g_{\psi h_1}~\epsilon_{\mu\nu\rho\sigma} \epsilon^{\rho\sigma\alpha\beta} \epsilon^\mu \epsilon_1^\nu q_{\beta}
\\ \nonumber && \times~{ 1 \over q^2 - m_{h_1}^2 + {\rm i} m_{h_1} \Gamma_{h_1} }~\left( g_{\alpha\gamma} - {q_{\alpha} q_{\gamma} \over m_{h_1}^2} \right)~p_4^\gamma\, .
\end{eqnarray}
This amplitude can be used to evaluate its decay width:
\begin{eqnarray}
&& \Gamma \left( X \to J/\psi h_1 \rightarrow J/\psi \pi^+ \pi^- \pi^0 \right)
\\ \nonumber &=& {g_{h_13\pi}^2\over3}{ (2\pi)^4 \over 2 m_X} \int {{\rm d}^3p_1\over(2\pi)^3 2E_1} {{\rm d}^3p_2\over(2\pi)^3 2E_2} {4\pi p_{3x}^2\over(2\pi)^6 2E_3 2E_4}
\\ \nonumber && \times \left|{1 \over {p_{3x} \over E_3} + {p_{1x} + p_{2x} + p_{3x} \over E_4} }\right| ~ \left|{1 \over q^2 - m_{h_1}^2 + {\rm i} m_{h_1} \Gamma_{h_1} }\right|^2
\\ \nonumber && \times~g_{\psi h_1}~\epsilon_{\mu\nu\rho\sigma} \epsilon^{\rho\sigma\alpha\beta} q_{\beta}
~ g_{\psi h_1}^*~\epsilon_{\mu^\prime\nu^\prime\rho^\prime\sigma^\prime} \epsilon^{\rho^\prime\sigma^\prime\alpha^\prime\beta^\prime} q_{\beta^\prime}
\\ \nonumber && \times \left( g^{\mu\mu^\prime} - {p^{\mu} p^{\mu^\prime} \over m_X^2} \right) ~ \left( g^{\nu\nu^\prime} - {p_{1}^\nu p_{1}^{\nu^\prime} \over m_{J/\psi}^2} \right)
\\ \nonumber && \times \left( g_{\alpha\gamma} - {q_{\alpha} q_{\gamma} \over m_{h_1}^2} \right) ~ \left( g_{\alpha^\prime\gamma^\prime} - {q_{\alpha^\prime} q_{\gamma^\prime} \over m_{h_1}^2} \right)~p_4^\gamma p_4^{\gamma^\prime} \, .
\end{eqnarray}
Again, the phase space integration is done in the reference frame where $p_3 = \left( E_3, p_{3x}, 0, 0 \right)$, and $p_{3x}$ satisfies $p_{3x}>0$ as well as
\begin{equation}
E_1 + E_2 + E_3 + E_4 = m_X \, .
\end{equation}

\section{Uncertainties due to the phase angle}
\label{app:phase}

\begin{table*}[hbt]
\begin{center}
\renewcommand{\arraystretch}{1.5}
\caption{Relative branching ratios of the $X(3872)$ evaluated through the Fierz rearrangement. In this table we fix the phase angle $\theta$ between the $S$- and $D$-wave coupling constants, $g^S_{D \bar D^{*}}$ and $g^D_{D \bar D^{*}}$, to be $\theta = \pi$.}
\begin{tabular}{ c | c | c | c | c | c | c}
\hline\hline
\multirow{2}{*}{Channels} & \multicolumn{3}{c|}{$|0_{qc}1_{\bar q \bar c}; 1^{++} \rangle$} & \multicolumn{3}{c}{$|D \bar D^*; 1^{++} \rangle$}
\\ \cline{2-7} & $I = 0/\theta_1^\prime = 0^{\rm o}$ &  $\theta_1^\prime = +15^{\rm o}$  &  $\theta_1^\prime = -15^{\rm o}$ & $I = 0/\theta_2^\prime = 0^{\rm o}$ &  $\theta_2^\prime = +15^{\rm o}$  &  $\theta_2^\prime = -15^{\rm o}$
\\ \hline\hline
${\mathcal{B}(X \rightarrow \eta_c f_0 \rightarrow \eta_c \pi \pi) \over \mathcal{B}(X \rightarrow J/\psi \omega \rightarrow J/\psi \pi \pi \pi)}$
& ~$\sim0.091$~ & ~$\sim0.091$~ & ~$\sim0.091$~ & ~$\sim0.091$~ & ~$\sim0.091$~ & ~$\sim0.091$~
\\ \hline
${\mathcal{B}(X \rightarrow \chi_{c1} f_0 \rightarrow \chi_{c1} \pi \pi) \over \mathcal{B}(X \rightarrow J/\psi \omega \rightarrow J/\psi \pi \pi \pi)}$
& ~$\sim0.086$~ & ~$\sim0.086$~ & ~$\sim0.086$~ & ~$\sim0.086$~ & ~$\sim0.086$~ & ~$\sim0.086$~
\\ \hline
${\mathcal{B}(X \rightarrow J/\psi h_1 \rightarrow J/\psi \pi \pi \pi) \over \mathcal{B}(X \rightarrow J/\psi \omega \rightarrow J/\psi \pi \pi \pi)}$
& ~$1.4 \times 10^{-3}$~ & ~$1.4 \times 10^{-3}$~ & ~$1.4 \times 10^{-3}$~ & ~$1.4 \times 10^{-3}$~ & ~$1.4 \times 10^{-3}$~ & ~$1.4 \times 10^{-3}$~
\\ \hline
${\mathcal{B}(X \rightarrow \chi_{c0} \pi) \over \mathcal{B}(X \rightarrow J/\psi \rho \rightarrow J/\psi \pi \pi)}$
& -- &  $0.024$  & $0.024$ & -- &  $0.024$  & $0.024$
\\ \hline
${\mathcal{B}(X \to J/\psi \omega \to J/\psi \pi \pi \pi) \over \mathcal{B}(X \rightarrow J/\psi \rho \rightarrow J/\psi \pi \pi)}$
& -- &  $1.6$~(input)  & $1.6$~(input) & -- &  $1.6$~(input)  & $1.6$~(input)
\\ \hline \hline
${\mathcal{B}(X \rightarrow D^0 \bar D^{*0} + D^{*0} \bar D^0 \rightarrow D^0 \bar D^0 \pi^0) \over \mathcal{B}(X \rightarrow J/\psi \omega \rightarrow J/\psi \pi \pi \pi)}$
& $0.021~t_1$ &  $0.034~t_1$ & $0.011~t_1$
& $4.5~t_2$ & $7.4~t_2$ & $2.4~t_2$
\\ \hline\hline
\end{tabular}
\label{tab:relative2}
\end{center}
\end{table*}

There are two different effective Lagrangians for the $X(3872)$ decay into the $D \bar D^*$ final state, as given in Eqs.~(\ref{lag:DDsS}) and (\ref{lag:DDsD}):
\begin{eqnarray}
\mathcal{L}^S_{D \bar D^{*}} &=& g^S_{D \bar D^{*}}~X^{\mu}~D^0~\bar D^{*0}_{\mu} + \cdots \, ,
\\ \mathcal{L}^D_{D \bar D^{*}} &=& g^D_{D \bar D^{*}} \times \left( g^{\mu\sigma}g^{\nu\rho} - g^{\mu\nu}g^{\rho\sigma} \right)
\\ \nonumber && ~~~~~ \times X^{\mu}~\partial_\rho D^0~\partial_\sigma \bar D^{*0}_{\nu} + \cdots \, .
\end{eqnarray}
There can be a phase angle $\theta$ between $g^S_{D \bar D^{*}}$ and $g^D_{D \bar D^{*}}$, which parameter is not fixed. We rotate it to be $\phi = \pi$, and redo the previous calculations. The results are summarized in Table~\ref{tab:relative2}, where only the relative branching ratio ${\mathcal{B}(|D \bar D^*; 1^{++} \rangle \rightarrow D^0 \bar D^{*0} (\rightarrow D^0 \bar D^{0} \pi^0)) \over \mathcal{B}(|D \bar D^*; 1^{++} \rangle \rightarrow J/\psi\omega (\rightarrow \pi \pi \pi))}$ is influenced by this parameter.
\begin{widetext}
Using the mixing angle $\theta^\prime_1 = +15^{\rm o}$, we obtain
\begin{eqnarray}
\nonumber && {\mathcal{B}\left(|0_{qc}1_{\bar q \bar c}; 1^{++} \rangle \rightarrow
 J/\psi\omega (\rightarrow \pi \pi \pi)
: ~\,J/\psi\rho (\rightarrow \pi \pi)~\,
: ~\,\chi_{c0}\pi~
: \eta_c f_0 (\rightarrow \pi \pi)
: \chi_{c1} f_0 (\rightarrow \pi \pi)
: D^0 \bar D^{*0} (\rightarrow D^0 \bar D^{0} \pi^0)~
\right) \over \mathcal{B}(|0_{qc}1_{\bar q \bar c}; 1^{++} \rangle \rightarrow J/\psi\omega (\rightarrow \pi \pi \pi)}
\\ &\sim&
~~~~~~~~~~~~~~~~~~~~~~~~~~~~~~~~~~~1~~~~~~~~ : ~~0.63~({\rm input})~~ : ~0.015~ : ~~0.091~(?)\,~ : ~~~0.086~(?)\,~~ : ~~~~~~~0.034~t_1~ \, ,
\end{eqnarray}
while using the mixing angle $\theta^\prime_1 = -15^{\rm o}$, we obtain
\begin{eqnarray}
\nonumber && {\mathcal{B}\left(|0_{qc}1_{\bar q \bar c}; 1^{++} \rangle \rightarrow
 J/\psi\omega (\rightarrow \pi \pi \pi)
: ~\,J/\psi\rho (\rightarrow \pi \pi)~\,
: ~\,\chi_{c0}\pi~
: \eta_c f_0 (\rightarrow \pi \pi)
: \chi_{c1} f_0 (\rightarrow \pi \pi)
: D^0 \bar D^{*0} (\rightarrow D^0 \bar D^{0} \pi^0)~
\right) \over \mathcal{B}\left(|0_{qc}1_{\bar q \bar c}; 1^{++} \rangle \rightarrow J/\psi\omega (\rightarrow \pi \pi \pi)\right)}
\\ &\sim&
~~~~~~~~~~~~~~~~~~~~~~~~~~~~~~~~~~~1~~~~~~~~ : ~~0.63~({\rm input})~~ : ~0.015~ : ~~0.091~(?)\,~ : ~~~0.086~(?)\,~~ : ~~~~~~~0.011~t_1~ \, .
\end{eqnarray}
\end{widetext}

\end{document}